\documentclass{aastex631}

\usepackage[normalem]{ulem}
\usepackage{textcomp}
\usepackage{gensymb}
\usepackage{subfiles}

\newcommand{\zmag}{mag$_{\mathrm{z}}$}

\shorttitle{Mining for strong gravitational lenses with self-supervised learning}
\shortauthors{Stein et al.}

\graphicspath{{./}{figures/}}

\begin{document}

\title{Mining for strong gravitational lenses with self-supervised learning}

\author[0000-0002-5193-516X]{George Stein}
\email{gstein@berkeley.edu}
\affiliation{Berkeley Center For Cosmological Physics, University of California, Berkeley, Berkeley, CA 94720}
\affiliation{Lawrence Berkeley National Laboratory, Berkeley, CA 94720}

\author[0000-0003-1142-3095]{Jacqueline Blaum}
\email{jrblaum@berkeley.edu}
\affiliation{Department of Astronomy, University of California, Berkeley, Berkeley, CA 94720}
\affiliation{Lawrence Berkeley National Laboratory, Berkeley, CA 94720}

\author{Peter Harrington}
\email{pharrington@lbl.gov}
\affiliation{Lawrence Berkeley National Laboratory, Berkeley, CA 94720}

\author{Tomislav Medan}
\email{tmedan@lbl.gov}
\affiliation{Lawrence Berkeley National Laboratory, Berkeley, CA 94720}

\author{Zarija Luki\'{c}}
\email{zarija@lbl.gov}
\affiliation{Lawrence Berkeley National Laboratory, Berkeley, CA 94720}

\begin{abstract}
We employ self-supervised representation learning to distill information from 76 million galaxy images from the Dark Energy Spectroscopic Instrument Legacy Imaging Surveys' Data Release 9. Targeting the identification of new strong gravitational lens candidates, we first create a rapid similarity search tool to discover new strong lenses given only a single labelled example. We then show how training a simple linear classifier on the self-supervised representations, requiring only a few minutes on a CPU, can automatically classify strong lenses with great efficiency. We present 1192 new strong lens candidates that we identified through a brief visual identification campaign, and release an interactive web-based similarity search tool and the top network predictions to facilitate crowd-sourcing rapid discovery of additional strong gravitational lenses and other rare objects: \href{https://github.com/georgestein/ssl-legacysurvey}{github.com/georgestein/ssl-legacysurvey}. 
\end{abstract}

\section{Introduction} \label{sec:intro}

The identification of rare objects is essential to facilitate a number of studies in astronomy. In recent years, the rapid advance of digital sky surveys has increased the size and complexity of data at an ever-increasing pace, giving researchers more opportunities to discover and analyze unique objects. However, this opportunity comes at the cost of an enormous amount of data to search through, such that relying on manual inspection of imagery by experts has long been rendered obsolete. Crowd-sourced classification campaigns such as the Galaxy Zoo projects \citep{GZ1, GZ2} allow for the visual inspection of a relatively large number of galaxies ($\mathcal{O}(10^5)$ -- $\mathcal{O}(10^6)$), but this number still pales in comparison to the full extent of current and incoming surveys. For example, the latest data release of the Dark Energy Spectroscopic Instrument (DESI) Legacy Imaging Surveys \citep{DECaLS} includes $\mathcal{O}(10^9)$ galaxies - a number far out of reach of even the most ambitious crowd-sourcing effort.

While visual inspection of the full dataset to find rare objects is unattainable, the recent success of deep convolutional neural networks (CNNs; \cite{DL}) for a wide variety of supervised classification and regression tasks in computer vision has been shown to be invaluable in conjunction with crowd-sourced astronomical efforts. Given a sufficient sample of human-labelled galaxies, such networks can be trained in a supervised setting to classify images with remarkable accuracy, which has led to their widespread adoption for sky survey science \citep{GZ3, Vega-Ferrero_morphology, HuangI}. Although powerful, this approach is not without issues, as it directly relies on the quantity and quality of the (manually assigned) labels, which tend to be from ``more interesting'' galaxies. As such, these labels are biased toward bright and large objects and generally targeted toward more common galaxy morphologies. In particular, supervised CNN classification can prove difficult when the number of known objects of a particular class is very small, as the quality of the supervised network degrades considerably without a sufficient number of labelled samples to learn from \citep{Hayat_2021}.    

One important class of rare objects is strongly gravitationally lensed galaxy systems, in which light from background sources is lensed by an intervening foreground galaxy or galaxy cluster, magnifying and distorting the appearance of the background source. As the lensing is a function of gravitational potential, its strength is dominated by the dark matter content of the lens and thus can be used to study the dark matter distribution of galaxies \citep{flux_ratio}. Additionally, the magnification of background sources allows for the study of high-redshift galaxy features that would otherwise fall below the detection threshold of astronomical instruments \citep{faint_lens}. When multiple images of the same background source are observed, the time delay of unique features traversing different path lengths can provide constraints on the expansion of the universe \citep{Holicow}. We refer the reader to the \citealt{strong_lensing_summary} for a comprehensive review of the subject.

As a result of the scientific value of strong lenses, a large targeted effort has been undertaken to search for them in existing sky survey imaging data sets. The primary challenge in doing so is their scarcity, as we expect only about one in $\mathcal{O}(10^4)$ galaxies to be strongly lensed \citep{Collett}. In recent years, the search for strong lenses has been dominated by the use of CNNs trained in a standard supervised learning setup \citep[e.g.][]{JacobsI, JacobsII, JacobsIII, Lanusse, PetrilloI, PetrilloII, LI, He_lens,  CanamerasI, CanamerasII, HuangI, HuangII, Fabrizio}\footnote{see the lens finding challenge \citep{strong_lens_challenge} or \citealt{ml-in-cosmo} for additional examples.}. After assembling the (small) sample of already known strong lenses (or a set of simulated strong lenses) and selecting a number of galaxy images that are not lensed, the CNN is trained in a binary classification setting on the training set until the performance on a holdout validation set is optimized. These studies have succeeded in discovering a number of new strong lens systems, but even after compiling all known lenses to date across all surveys, the number of available training samples is a major limiting factor in achieving accurate predictions with this approach.

The lack of quality training labels coupled with the enormous quantity of data coming from current and near-future sky surveys motivates the need for scalable approaches capable of {\textit{learning from unlabeled data}}. Rather than training specialized networks for each individual classification or regression task, learning without labels allows for more generalized models capable of performing multiple tasks in parallel, while simultaneously benefiting from access to orders of magnitude more data. A preliminary application of unsupervised methods to lens identification was done by \citealt{stronglens_AE}, who applied a convolutional autoencoder to simulated data and clustered the resulting features to classify lenses. An exciting alternative to this purely unsupervised feature learning is the emerging paradigm of self-supervised learning, which aims to distill or extract useful information from images without requiring supervision labels for each sample in the dataset \citep{Zhai_2019_ICCV}. Self-supervised methods can make use of very large unlabeled datasets \citep[e.g., ][]{SEER}, which would be cumbersome or otherwise impossible to manually label, and build meaningful embeddings or representations by solving contrived tasks during training that require some high-level understanding of the input features. The representations produced by self-supervised models can then be directly used or fine-tuned for machine learning tasks of interest, and this approach is able to outperform standard supervised machine learning models, particularly when the number of data labels available for supervision is low \citep{chen2020simple, chen2020big, chen2020improved}. Compared to simpler unsupervised feature learners like auto-encoders, self-supervised models are known to produce higher-quality representations with far more predictive power \citep{bigan, bigan2} that generalize well to many different tasks \citep{DINO}.

The utility of self-supervised models applied to astronomical imagery was recently showcased by \cite{Hayat_2021}, who found that self-supervised pretraining on a large unlabeled set of SDSS galaxy images improves performance on tasks like redshift estimation and morphology classification, and that these performance gains are most significant when the number of labels for supervised training is limited. \cite{Hayat_2021} also showed that the representation space learned in self-supervised pretraining is semantically meaningful and readily provides a similarity metric that can identify additional examples of a query object, such as an observational error or anomalous galaxy. As automated strong lens detection is a problem inherently limited by the number of labeled examples, self-supervised models are thus an exciting prospect for quickly identifying new candidates given a large set of galaxy imagery.

In this work, we apply self-supervised learning to Data Release 9 (DR9) of the DESI Legacy Survey \citep{DECaLS} to conduct a search for strong gravitational lenses. Strong gravitational lensing can result in a number of different visual configurations, including multiple appearances of the same background galaxy or extended arcs curving toward the  central source, but determining whether the configuration is a true gravitational lens by visual inspection can be difficult. Therefore, unless the system is spectroscopically confirmed (or, in the case of faint lens features, observed at higher resolutions using other instruments), we often cannot use images to conclusively determine whether or not a system is truly a gravitational lens. As such, in this work, we consider {\textit{strong lens candidates}}: images that contain characteristics of strong gravitational lensing as viewed in the DESI Legacy Survey imaging. In Section~\ref{sec:data} we first give an overview of the images and strong lens labels used, followed by a description of our self-supervised model and supervised classifiers in Section~\ref{sec:methods}. In Section~\ref{sec:results} we detail our discovery of 1192 new strong gravitational lens candidates and present our publicly available similarity search tool to facilitate the discovery of additional gravitational lenses beyond those presented here.  Lastly, we conclude in Section~\ref{sec:conclusions}.

\section{Dataset}
\label{sec:data}
\begin{figure*}
    \centering
    \includegraphics[width=1.0\textwidth]{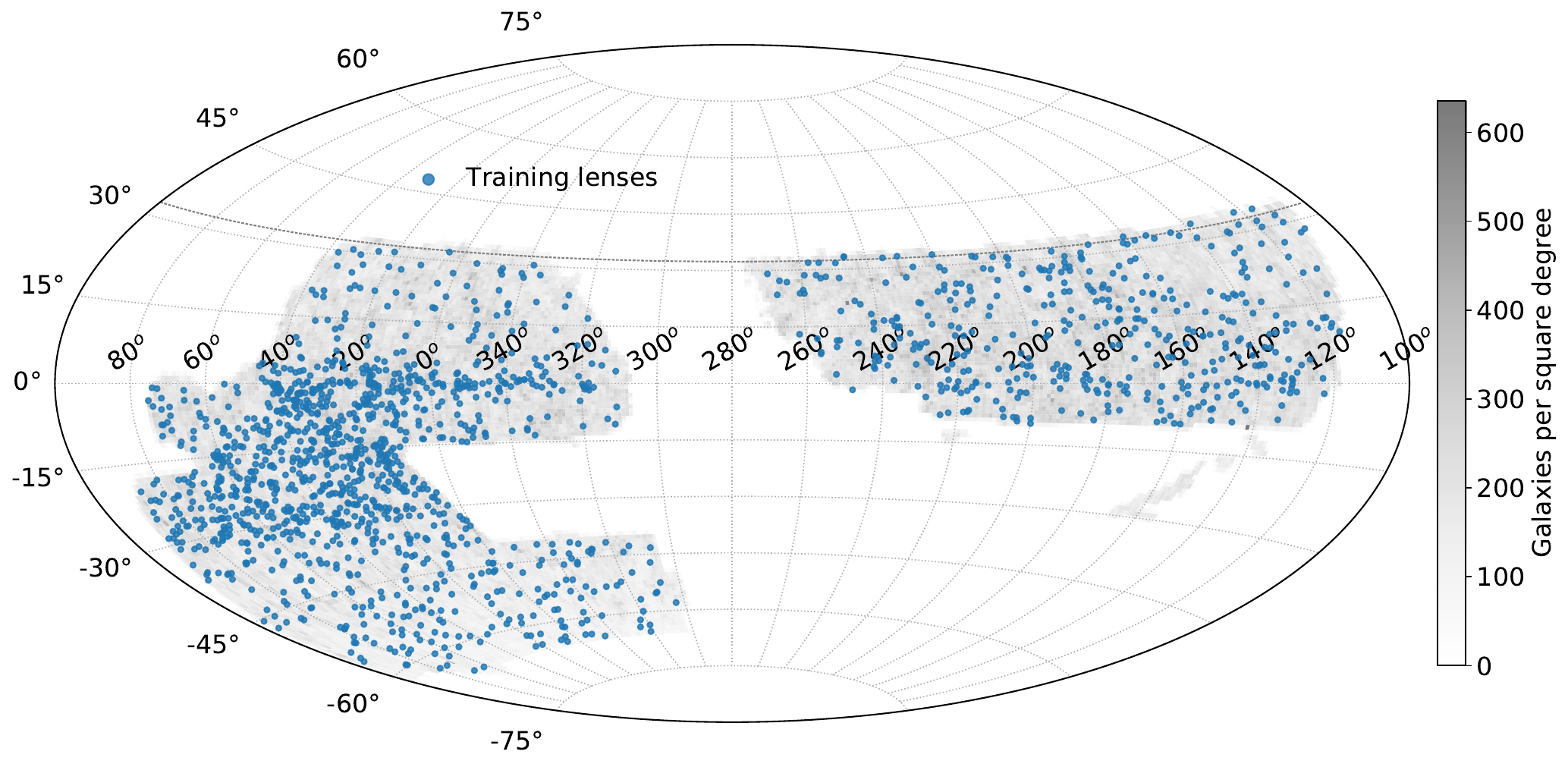}
    \caption{Right ascension and declination of previously identified DECaLS strong lens candidates, with the background shaded by the number of the 3,500,000 galaxies used to train the self-supervised network in each square-degree region. The gray line at $\delta=32.375$ denotes the split between the north and south surveys.}
    \label{fig:sky_plot_data}
\end{figure*}

\subsection{DESI Legacy Survey}

The DESI Legacy Imaging Surveys\footnote{\href{https://www.legacysurvey.org/}{legacysurvey.org/}} \citep{DECaLS} cover approximately 19,721 square degrees of extragalactic sky visible from the northern hemisphere in three optical bands (g, r, z).  The survey footprint is split into two components, north and south. The observations in the north region, $\delta > 32.375$, were undertaken jointly by the Beijing-Arizona Sky Survey (BASS; g and r bands) and the Mayall z-band Legacy Survey (MzLS; z-band), while the southern part of the survey, $\delta < 32.375$, uses data from the DECam Legacy Survey (DECaLS). The DECaLS program made use of other DECam data, the most significant of which is from the Dark Energy Survey (DES), which includes a 5000 deg$^2$ contiguous area in the south Galactic cap. DECaLS explicitly did not reimage the DES area and instead incorporated the DES imaging itself. We use the latest data release, DR9, released in January of 2021. The DR9 does not contain significant new observations but builds on DR8 by improving the reduction techniques and procedures used for the Legacy Surveys. Imaging from the Legacy Surveys is first reduced through the NOIRLab Community Pipeline\footnote{\href{https://www.noao.edu/noao/staff/fvaldes/CPDocPrelim/PL201_3.html}{noao.edu/noao/staff/fvaldes/CPDocPrelim/PL201\_3.html}} before being processed using the Tractor\footnote{\href{https://github.com/dstndstn/tractor}{github.com/dstndstn/tractor}}.  The Tractor identifies unique sources in the imaging survey, measures a large number of photometric features, and fits for a number of quantities, including morphological type.

Our self-supervised approach allows us to learn from the entire dataset and does not require restrictive data cuts to narrow down objects to those most likely to show features of strong lensing, such as color or magnitude cuts. Rather, we aim to use every galaxy in the DESI Legacy Imaging Surveys. We keep every source in the sweep catalogues (a subset of the information about each source from the full Tractor catalogues) that was not identified as a star, i.e., did not have a best-fit morphological model of point-spread function (PSF)=stellar. For both the north and south surveys, we compile an initial list of all non-PSF sources and keep those with a z-band magnitude (denoted \zmag) less than 20. This cut was introduced to keep the data at a more manageable size and to focus on galaxies that have been included in previous lens searches \citep{HuangII}. For the south survey, we also include the next 20 million sources with \zmag $> 20$, sorted by decreasing brightness. 

For each source in the list we extracted a 256$\times$256 pixel cutout in each of the (g, r, z) bands at 0.262 arcsecond resolution, centered at the right ascension and declination of the source. A very small fraction of galaxies that existed in the sweep catalogues resulted in an error when requesting the cutout, so these sources were excluded from our final catalogues. In total, this resulted in 14,174,203 images from the north survey and 62,272,646 images from the south, 42,272,646 with a \zmag $< 20$, for a total of 76,446,849 images. Sources in, e.g., galaxy clusters can be separated by less than the angular extent of an image, so separate images of nearby sources will include overlapping regions of the sky. The vast majority of galaxies and strong lenses cover angular extents much less than the full 256 pixel cutout, so we choose to train all networks on 96$\times$96 pixel (25$\times$25 arcsecond) crops. In order to perform the required set of data augmentations during training, mainly random rotations and jitter, we save 152$\times$152 pixel versions to disk, which results in a dataset size of 20TB. Files are saved in hd5f format and split into arrays of 1 million images (i.e. each file contains an image array of size (1,000,000, 3, 152, 152) and the corresponding sweep catalogue information for each source).

\subsection{Compilation of previously found lenses}
\label{sec:data_lenses}

We compiled a list of strong lens candidates from previous lens-finding campaigns, using all candidates from the following sources:
\begin{itemize}
    \item The Master Lens Database\footnote{\url{http://admin.masterlens.org/index.php}} as of late July 2021.
    \item \citet{HuangI, HuangII} with tables available on their project website\footnote{\url{https://sites.google.com/usfca.edu/neuralens}}
    \item A number of previous works with tables published through the VizieR database\footnote{\url{https://vizier.u-strasbg.fr/viz-bin/VizieR}} \citep{CanamerasI, JacobsI, JacobsII, JacobsIII, Diehl, Carrasco, Pourrahmani}
    \item The Survey of Gravitationally-lensed Objects in HSC Imaging (SuGOHI) Candidate List\footnote{\url{http://www-utap.phys.s.u-tokyo.ac.jp/~oguri/sugohi/}} \citep{SUGOHI_I,SUGOHI_II,SUGOHI_III,SUGOHI_IV,SUGOHI_V,SUGOHI_VI}. 
\end{itemize}
Some of the above works overlap with the Master Lens Database, but we included their data tables separately to ensure all sources were included. We disregard the ``quality'' labels indicating the respective team's level of confidence that each candidate is actually a lens, as the classification criteria for these labels are inconsistent throughout the literature. During publication of this paper, after the conclusion of our lens search, \citet{rojas2021strong} performed an additional supervised CNN search on the DES region, finding 186 new lens candidates, and \citet{LIII} presented 97 new high-quality strong lensing candidates. We also became aware of another separate repository of known lenses\footnote{\url{https://www.astro.rug.nl/lensesinkids/}}, that were already included in our analysis through the Master Lens Database.  

To first determine which of the lens candidates overlap with our image data set, we cross-match the right ascension and declination from our lens candidate list with the approximately 76 million sources in our DR9 sample, and define a match as any lens that has a DR9 source separated by less than the size of our images (96 pixels). For galaxy clusters with multiple sources near the given lens position we flag the nearest source as ``the lens'' regardless of whether it is in the center of the cluster. We additionally flag any DR9 sources that reside near a lens, but are not the closest, by noting their distance from the lens. To prevent any ambiguity during training of our networks in Section~\ref{sec:classification} we make sure not to sample sources that happen to lie nearby any labelled lenses, as the lens features may still appear in the image.

This procedure results in nearly 6000 lens candidates that overlap with sources in DR9, but by our own lens classification criteria the majority do not appear to have clear lensing signatures in DR9 imaging\footnote{We examine images after converting from nanomaggies to RGB values using the same transformation as the legacy survey viewer}. This discrepancy is potentially due to limitations in the angular resolution or noise levels of DR9, or due to unclear classifications of what constitutes a lens candidate in the variety of sources compiled. As we explain in the following section, we train all models only on south sources with \zmag $<20$. In this subset, we visually inspect the lens candidates and keep only the ones that, in our opinion, exhibit distinguishable lensing features in DR9 images. We do not include images with very faint or ambiguous features. We refer to this final catalogue of 1615 lenses as {\textit{DECaLS strong lenses}}, and use only this sample for training and validating our supervised networks.

We show the distribution of lenses on the sky in Figure~\ref{fig:sky_plot_data}. As expected, we see a higher density of lenses in the DES region due to the greater imaging depth and its heightened focus by previous lens searches. In addition, given our highly subjective visual selection of often-ambiguous images (which is based on an already-biased sample of previously discovered lenses), our training set has likely over- or undersampled certain lens classes and configurations, which affects the set of newly uncovered lens candidates. Thus, the lens-candidates we present in this work should not be considered as a representative sample of all those remaining to be found in sky survey datasets, but rather as a complex combination of the known sample of lenses available for training, the biases introduced by the specific subset of humans who examined them for this work, and the inherent biases of classification with highly nonlinear models in a class-imbalanced problem. Rather than assembling a training set of lenses by visual inspection, one could instead learn from a set of simulated lenses \citep{JacobsI, LI}, although this procedure does not come without its own set of biases, as the ``true'' distribution of lenses is still unknown, and the lensing profiles used in the simulations may not perfectly reflect reality. As we are working from a data-driven perspective and have developed a tool to find lenses given a single example, we chose to opt for the set of human-classified lenses as they appear in our imaging in the hope that they include more outlying examples from those more commonly exhibited in lensing systems. However, we note that a set of simulated lenses would allow for very similar investigations.

\section{Methods}
\label{sec:methods}

\begin{figure*}
    \centering
    \includegraphics[width=1.0\textwidth]{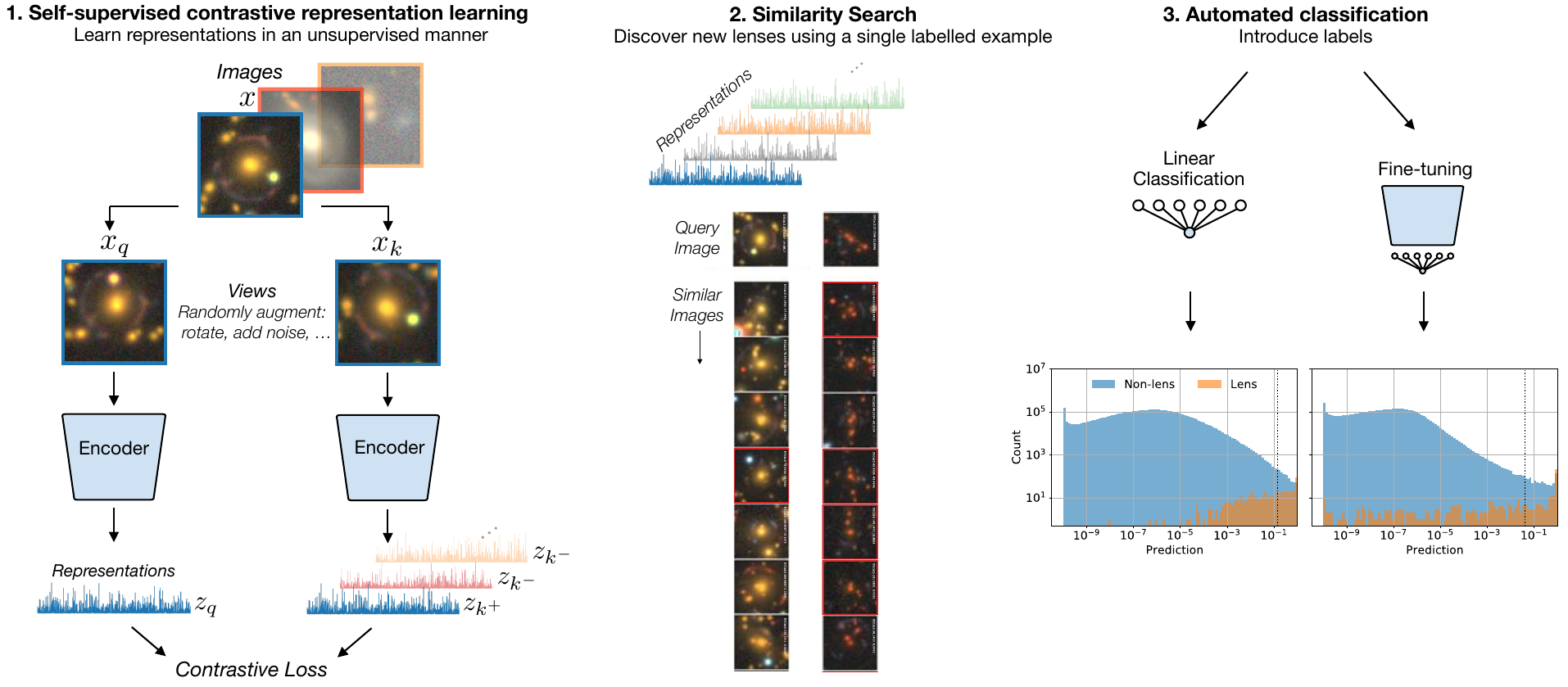}
    \caption{Illustration of self-supervised training (left; see \cite{Hayat_2021}) and its use cases for discovering objects through similarity search (middle) and automated classification (right).}
    \label{fig:schematic}
\end{figure*}

In this section, we describe the machine learning methods used to sift through our galaxy catalogs and detect new strong lens candidates. We provide a schematic of our workflow in Figure \ref{fig:schematic}. The first step in our process is to train a self-supervised model on a large body of unlabeled data, allowing the network to see many types of galaxies and learn from the natural distribution of morphologies and features within the survey sample. Then, we can take the representations learned by the self-supervised model and use them to identify new lenses via similarity searches and automated classification.

\subsection{Self-supervised Pretraining on Unlabeled Images}

Modern self-supervised learning has been shown to distill highly information-rich representations from standard machine learning datasets without utilizing any labelled information, with the classification performance of a simple linear classifier trained on such representations rivaling the performance of fully supervised CNN models \citep{chen2020simple, he2020momentum,  BYOL}.\footnote{See self-supervised ImageNet  \href{https://paperswithcode.com/sota/self-supervised-image-classification-on}{linear classification accuracy} by publication date} Such datasets contain a ground-truth label for every image, allowing for robust measurements of the quality of the learned representations, as better representations will lead to increased performance on downstream supervised tasks using these labels. However, for many scientific datasets, especially sky surveys, the number of ground-truth labels is minimal compared to the number of images, and the labels themselves often have nonnegligible noise or bias. Thus, assessing representation quality via robust downstream supervised evaluations is much more challenging, so tuning hyperparameters and choosing data augmentations for the self-supervised model can be difficult.

Rather than supervised evaluation, \citet{ssl_representation_quality} found that the evaluation on an additional self-supervised task (e.g., rotating images and predicting the rotation from the representations) was strongly correlated with the supervised evaluations and could be used in place of supervised labels to determine representation quality. Unfortunately, their strongest results were for a rotation evaluation task, which is incompatible with the rotation-invariant representations of galaxies we desire in this study (as there is no preferential orientation of galaxies on the sky). A jigsaw task \citep{ssl_jigsaw} may prove useful for our dataset, but we leave this for future work and instead evaluate the quality of our self-supervised models according to their linear classification performance on our validation set of strong lenses.

To limit the amount of hyperparameter tuning, we closely follow \cite{Hayat_2021} in designing the architecture and training procedure for our self-supervised model. In this setting, the backbone of the model is a CNN encoder that takes an image $\bf{x}$ as input and produces a lower-dimensional representation $\bf{z}$. The encoder learns to make meaningful representations by associating augmented views of the same image as similar and views of different images as dissimilar via a contrastive loss function. The representations for different images are maintained in a queue during training, which extends the number of contrasting examples available to the model at each training step beyond just those available in a given minibatch. More details on this approach can be found in \cite{Hayat_2021} and \cite{chen2020improved}. We use the same ResNet50 network and training hyperparameters as \cite{Hayat_2021}, but increase the queue length to $K=262,144$ to accommodate our larger training set. As our encoder architecture is a standard ResNet50, the number and size of convolutional kernels and pooling layers results in a representation vector of 2048 dimensions. This is strictly a result of the encoder architecture used, and while a representation of this size has been shown to perform well on downstream tasks on common curated industry datasets, it is an open question as to whether it is an optimal size for sky survey datasets, which often exhibit a smaller degree of image diversity.

The key to successful self-supervised pretraining is to produce the differing views of each image using a carefully crafted set of image augmentations that reflect symmetries, uncertainties, or noise in the dataset. These augmentations must realistically perturb the images, making it nontrivial for the model to associate augmented pairs of the same image while still preserving the important features in the input. We choose the following set of augmentations, applying each of them in succession to images during pretraining with the order listed below,
\begin{itemize}

\item \textit{Galactic extinction:} We first deredden the image according to its tabulated SFD $E(B-V)$ value \citep{SFD1998Dust}, simulating a view of the galaxy with no foreground dust. Then, we randomly sample a new $E(B-V)$ value from a lognormal distribution fit to the dataset and artificially redden the image with this random $E(B - V)$. Details on the lognormal $E(B - V)$ distribution and photometric conversions between $E(B-V)$ and transmission can be found in Appendix \ref{app:augmentations}.

\item \textit{Rotation/Orientation:} We randomly flip the image across each axis with 50\% probability, then rotate by a random angle sampled from $\mathcal{U} (0, 2\pi)$.

\item \textit{Size Scaling:} We randomly resize the image to between 90\% and 110\% of its original size to simulate views of a galaxy at different distances. We do not account for the change in redshift or resolution corresponding to such changes in distance and simply rescale with bilinear interpolation.

\item \textit{Point Spread Function (PSF) blur:} We model additional PSF smoothing (on top of the existing inherent PSF smoothing in the image) by applying Gaussian blur to each channel. The amount of blurring (i.e., the width $\sigma$ of the blurring kernel) is parameterized by lognormal fits to the PSF distribution found in the data (see Appendix \ref{app:augmentations} for details).

\item \textit{Jitter and crop:} We translate the image center along each respective axis by $i,j$ pixels, where $i,j \sim \mathcal{U}(-7,7)$, before cropping out the central 96$\times$96 pixels.

\item \textit{Gaussian noise:} We add additional Gaussian noise to the image (on top of the inherent noise already present) by sampling a noise level from lognormal distributions tuned for each filter channel. The noise in each channel is uncorrelated, since images are taken at different times and/or with different telescopes. See Appendix \ref{app:augmentations} for details on how each channel's noise distribution is parameterized.

\end{itemize}

With these augmentations, we perform self-supervised pretraining on our training set, which was constructed with the following considerations.

A significant number of the full 76 million images are faint red galaxies with a spatial size in the image spanning no more than a few pixels. This dominant class of objects would introduce a high degree of degeneracy due to the similarity between samples, which would limit the effectiveness of the contrastive loss used in self-supervised training. Thus, to ensure a more diverse sample of galaxy images, we choose to train on sources with \zmag~$<20$ and to sample our training set equally in \zmag~rather than randomly sampling over the full dataset. The magnitude cut (aligning with that chosen by \citep{HuangI}) removes the dimmest galaxies which exhibit the highest degree of visual similarity, and the uniform \zmag sampling helps to ensure that the training set is not dominated by the the vastly more numerous dim galaxies. Although the brightest and most nearby galaxies will not constitute strong lenses we do not use any minimum magnitude cut to remove them from the self-supervised training set; the goal at this stage is only to learn task-agnostic galaxy representations from a diverse set of galaxy images that can later be used for a variety of different downstream tasks. We show the magnitude distribution of the training sample compared to the full sample in Appendix~\ref{app:augmentations}.

In addition to uniform magnitude sampling, we train only on images from the south region of the survey. The north and south surveys were observed with different telescopes and bandpasses; therefore, images from these regions have different statistical properties. South images have the best signal-to-noise ratio and angular resolution, and  lenses and sharp features are easier to distinguish. We posit that training using this higher-quality sample and then simply using the models on the lower-quality north sample will allow the network to learn a more information-rich representation than training on both surveys simultaneously. Careful crafting of region-specific augmentations could possibly allow for the combination of these two data samples in a self-supervised learning approach, but we leave this for future work.  

These efforts result in a curated subset of 3.5 million images, and we are able to train the model in just 8 hours on eight NVIDIA Tesla V100 GPUs. After the self-supervised pretraining phase, our CNN encoder was applied to all 76,446,849 images from both the south (DECaLS) and north (MzLS/BASS) regions of DR9 to distill features from each image into a representation vector $\mathbf{z}$ of length 2048. This procedure extrapolates beyond the training data to both objects of lower magnitude (south images with \zmag $>20$) and to images with inherently different statistical properties (north images), yet we will show that the models nevertheless exhibit strong generalization properties outside of the distribution seen during training. We believe this generalizability simply stems from the high degree of image similarity between \zmag~$=20$ objects and those slightly dimmer than the cutoff and from the overall similarity of the north images compared to the south. The north images exhibit a higher level of noise and slightly different colors (specifically in the z band), and they have the same pixel size and the same range of morphologies, features, and shapes as those in the training set from the south. Although subtle selection effects are possibly introduced by applying our models outside of the training set, we find a significant number of lenses in these regions, thus showing that the models have useful generalization abilities for the applications of this paper.    

In the following sections, we describe how these representations are used to search the dataset for objects similar to known lenses, and then demonstrate their ability to directly reveal new lens candidates via automated classification.

\subsection{Image similarity search in representation space}

A powerful advantage of self-supervised pretraining is the ability to provide a measure of semantic similarity between any two images. Given the symmetries and noise inherent to the dataset, a notion of similarity is difficult to construct in pixel space, as noise levels and image rotations drastically change the individual pixel values while leaving the semantic information in the image unchanged. Rather than operating in the pixel space, the representations we distill from each image in the dataset retain the overall information, such as the size and relative orientation of the galaxies, the number of sources, the clustering, and the color, while removing the noise and symmetries described by the data augmentations \citep[for a visualization of how galaxy properties are organized in the representation space of a self-supervised model, see][]{Hayat_2021}. While we find that the representations contain information highly relevant to lens identification, it is important to note that the self-supervised pretraining is task agnostic, and {\textit{not a single lens label has been used during training}}. We can utilize the similarity search to uncover additional rare objects of any type, not only strong gravitational lenses.  

The contrastive loss in self-supervised pretraining encourages semantically similar images to have similar representations, so we can easily define a similarity metric in representation space (which is a lower dimensionality than the input images). We choose to measure similarity between any two images $\bf{x_i}$ and $\bf{x_j}$ by the cosine similarity (normalized scalar product) of their representation vectors $\bf{z_i}$ and $\bf{z_j}$:
\begin{equation}
    \text{similarity}(\bf{z_i},\bf{z_j}) = \bf{z_i} \cdot \bf{z_j}/ (\| \bf{z_i} \| \, \| \bf{z_j} \| ).
\end{equation}

For each of the 42 million images in the south region of DR9 with \zmag $< 20$ we calculate the 999 most similar images, and save the resulting ``similarity array'' (size $\sim$42 million $\times$ 1000) for later reuse. Computations are performed on eight GPUs using Facebook AI Research's Faiss\footnote{\url{https://github.com/facebookresearch/faiss}} library for efficient similarity search\footnote{While we calculated the cosine similarity without employing any of the compression techniques available in Faiss, we found that it still outperformed its NumPy equivalent on CPU and benefited from an additional speed-up on GPU}. To avoid unnecessary computations we did not compute the full $N\times N$ similarity matrix ($N=42$ million), but for each source we limited the search to sources within a magnitude range of 0.5 in r-band. We found that this resulted in nearly equivalent results as the full search, as the model naturally learned that brighter galaxies do not have significant similarity with dimmer ones. The pre-computed similarity array allows for rapid similarity searches of any image in the dataset; given a query image, the most similar images are immediately available to examine (see \cite{neurips_similarity} for a discussion of additional use cases for our similarity search tool).

\subsection{Automated classification of strong lenses}
\label{sec:classification}

We perform automated classification of strong lens candidates through two different supervised methods, both of which rely on a pre-trained self-supervised model. The first approach is to simply perform linear classification directly on the self-supervised representations, and the second is to fine-tune the self-supervised model for classification.

\subsubsection{Training and validation sets}
\label{sec:train_val_sets}
As described in Section~\ref{sec:data_lenses}, our efforts in compiling a list of known lenses from the literature and cross-matching them with our DR9 galaxy images yielded a total of 1615 well-defined lens images in the south region. We use a random selection of 70$\%$ (1130) of these lenses for training and save the remaining 30$\%$ (485) for validation. We do not train on any images from the north region or on sources with \zmag $> 20$, and only use these for lens searching. Unlike some previous supervised CNN lens identification works, we do not hand-select the nonlens samples to visually resemble different types of galaxy and cluster morphologies. Hand selection is labor intensive and can artificially inflate the occurrence of specific galaxy types in the training set as compared to the full dataset that we will be performing the lens search on. Instead, we simply select nonlens samples randomly from the full dataset, excluding those with a lensing galaxy near the central galaxy and those flagged as lenses in other data sets but hard to distinguish visually in DR9 imaging. This approach allows for greater coverage of the survey with minimal manual effort and allows us to build training and validation sets that actually reflect the scarcity of lensed galaxies ($\mathcal{O}(10^{-4})$) expected in the data. 

Our random selection of nonlenses from the true data distribution introduces a small amount of class leakage, where undiscovered lenses are mislabeled, but the effects of this leakage are marginal due to the rarity of lenses. To confirm this, we experimented with using an initial linear model to predict labels for the training and validation nonlenses and then visually examined the images with the highest predictions to identify and remove this contamination. However, the level of contamination was small ($\sim$40 of the 4 million ``nonlenses'' were later found to be lens candidates), and retraining the linear classification model with these more accurate labels returned highly similar predictions, so we kept the original training and validation labels.

To construct a realistic validation set reflecting the scarcity of strong lens populations, we selected 10,000 nonlenses for every lens. This differs from many strong lens searches in the literature, which instead sample validation sets using the same lens-to-nonlens ratios (denoted lens:nonlens) as they selected for their training set (e.g., 1:1 \citep{JacobsIII} or 1:33 \citep{HuangI}). Using a more equal ratio will significantly inflate the performance metrics calculated across these validation sets compared to the true data (1:10,000) when undertaking the search for new lenses. The optimal model as measured using a 1:1 validation set is not necessarily optimal on a true 1:10,000 population and is likely to flag a significant number of false positives. Thus, we constructed our validation set to instead represent a likely realistic fraction of lenses in the universe. We treat the lens:nonlens ratio in the training set as a hyperparameter and search over different settings to determine the optimal level of class imbalance. We present our findings for this search in section \ref{sec:lens_classification_results}.

\subsubsection{Model Evaluation}

By searching for rare objects, we are performing a classification task in a scenario with several specific characteristics. Namely, our dataset is highly class-imbalanced, and we have limited time to manually assess predicted lens candidates for quality, so we opt for slightly different performance metrics than those canonically used in classification. As our goal is to identify the greatest number of lenses in a limited amount of human time, we decided to restrict our time spent visually inspecting the top network predictions to a few hours. After performing an initial test to see how many images three of the authors could classify per unit time, we estimated our ``classification rate'' to be $\sim10,000$ images in a few hours. Thus, if our training set was an unbiased sample of the lenses in the universe, an optimal model would maximize the number of known lenses in the top $N=10^4$ predictions and minimize the number of false positives; for practical purposes we are unconcerned with the network performance on the rest of the data set. Our human classification procedure does not include a detailed examination of each image but rather a fast scan of 500 images laid out in a 5x100 grid, devoting approximately 0.5-1 second to each individual image. This procedure leads to a lens candidate selection biased toward more distinct lensing characteristics and configurations and likely misses a number of lower-quality candidates, but it allows for a rapid rate of lens finding. 

Based on these initial estimates, we use only the top $N=1139$ predictions on our validation set to evaluate model quality. This particular value of $N$ was chosen by scaling down the total number of images we expect to visually classify ($10^4$) to adjust for the difference in size between the validation set (4.85 million galaxies) and our full search size (which we initially scoped to contain $\sim$42.5 million images in the south survey). This procedure is in contrast to the standard approach of choosing a probability cutoff $c_p$ to manage the trade-off between precision and recall, as our ``cutoff'' $c_{\mathrm{top }N}$ is simply determined such that the number of samples above the cutoff is $N=1139$. Then we can easily asses model quality by computing the precision, 
\begin{equation}
    \label{eq:precision}
    \mathrm{Prec}_{\;( \mathrm{top} \; N)} = \frac{TP}{FP+TP},
\end{equation}
where $TP$ and $FP$ are the number of true positives (the number of true lenses) and false positives (the number of nonlenses) in the top $N$ predictions of the model, respectively. We note that the choice of $N=1139$ is of little consequence to the chosen model's utility - i.e. we found the model that maximizes the number of lenses in the top 1139 predictions will be the same model that maximizes the number of lenses in the top 2000 predictions. Thus, for consistency, we choose to fix $N=1139$ throughout the experiments of this paper regardless of the exact final search size or number of lenses in the final validation set.    

\subsubsection{Linear Classification}
\label{sec:linear_classification}

As shown in \citet{Hayat_2021} for classification of galaxy morphologies, linear classification on self-supervised representations of images achieves superior performance to full supervised training of a CNN when the amount of labelled data is small. This is despite the fact that the self-supervised representations have only been trained to group similar images together, and have not been explicitly trained to pick out properties of any specific objects like strong gravitational lenses. As we will demonstrate using a similarity search, unlabelled images that are predicted as a high probability of being a lens reside nearby the labelled samples (i.e., known lenses) in self-supervised representation space. Thus, the 2048-dimensional representation space can be partitioned into binary class predictions (nonlens and lens) by a simple linear classifier of the form $\hat{y} = {\bf{W}}^T {\bf{z}}  + b$, where $\hat{y}$ is the predicted lens probability, $\bf{z}$ is the representation vector, $\bf{W}$ is a 2048-dimensional vector of weights, and $b$ is a bias. We also experimented with classifying representations using a random forest classifier, but found this more complex nonlinear model to be prone to overfitting and all configurations tested performed worse on the validation set.

We train our linear classifiers in PyTorch~\citep{pytorch} using a binary cross-entropy loss function with LBFGS optimization. Our final model is the epoch that achieves the largest number of true positives in the top 1139 predictions on the validation set, as explained in the previous section. The simplicity of the model and the low dimensionality of the representations allows for training to proceed rapidly on a single CPU in a straightforward single-batch gradient descent, and training on the 4.5 million samples of the 1:4000 training set only requires 5-10 minutes on a CPU (45s per epoch), compared to many hours on multiple GPUs in a supervised CNN setup. This has the potential for a significant impact beyond the strong lens classifier application demonstrated here. Rather than requiring both the access to computational facilities to host the full image dataset and perform multi-GPU training, and the machine learning expertise to train complicated CNN models in parallel, self-supervised representation learning reduces the problem to one that can be solved with minimal computing resources and computational background, opening the door to crowd-sourced ``citizen science'' efforts like those employed for galaxy morphology studies (e.g.,  \cite{GZ1, GZ2}).

\subsubsection{Fine-tuned Self-supervised Classification}
\label{sec:finetuning_methods}

While linear classification on the self-supervised representations achieves high-quality classification results, we can instead attach a fully connected layer on the 2048-dimensional output of the self-supervised encoder CNN and fine-tune the network in a supervised setup using the available lens labels. By applying a smaller learning rate to update the CNN encoder parameters and a larger learning rate on the linear classification head we can train the linear classifier from a random initialization, while only marginally modifying the CNN parameters optimized during the self-supervised learning phase. This can improve the classification results beyond those achievable with linear classification on the original self-supervised representations, but the effect can be small in the few-label regime \citep{Hayat_2021}, and can overfit with a small number of labels for a highly class-imbalanced task. However, there is evidence that fine-tuning a self-supervised network improves model robustness over purely supervised learning \citep{ssl_robustness}.

Works focusing on supervised CNN classification of strong lenses utilize minibatch gradient descent due to memory constraints introduced by large CNN models and image dimensionality, with minibatches of size $\sim$128 images. Each minibatch requires instances of both classes to efficiently learn, and cannot be completely dominated by the negative class (nonlens). Therefore, to ensure that each minibatch on average has a few lenses either the class imbalance in the training set must be chosen to be much less than the true data (e.g. 1:1 \citep{JacobsIII} or 1:33 \citep{HuangI}), or oversampling techniques which sample lenses more often than their nonlens counterparts must be employed. Our linear classifier, trained with a single large batch, does not have these requirements and we can easily train with a high degree of class imbalance. To use the same training set as the linear classifier, we oversampled lenses during fine-tuning to ensure a nonzero average number of lenses in each batch. This oversampling factor is an additional hyperparameter of the training setup that we explore during training to optimize the model performance on the validation set.

Training was conducted in PyTorch using a binary cross-entropy loss. We augmented images at each training epoch with jitter/crop, random rotations, Gaussian noise, and Gaussian blur; thus, each sampled version of a given lens will have different realizations of the augmentations. We trained using eight NVIDIA Tesla V100 GPUs for 40 epochs with a batch size of 512 using the SGD optimizer. We implemented a learning rate of 0.01 for the classification head and 0.001 for the remaining network parameters as in \citet{Hayat_2021}. Training requires approximately 1 hour per epoch. Note that this does not take into account the compute time required for self-supervised pretraining, which does not need to be undertaken separately for each supervised hyperparameter setup.

\section{Results}
\label{sec:results}
\subsection{Identification of lens candidates via similarity search in representation space}

\begin{figure*}
    \centering
    \includegraphics[width=1.0\textwidth]{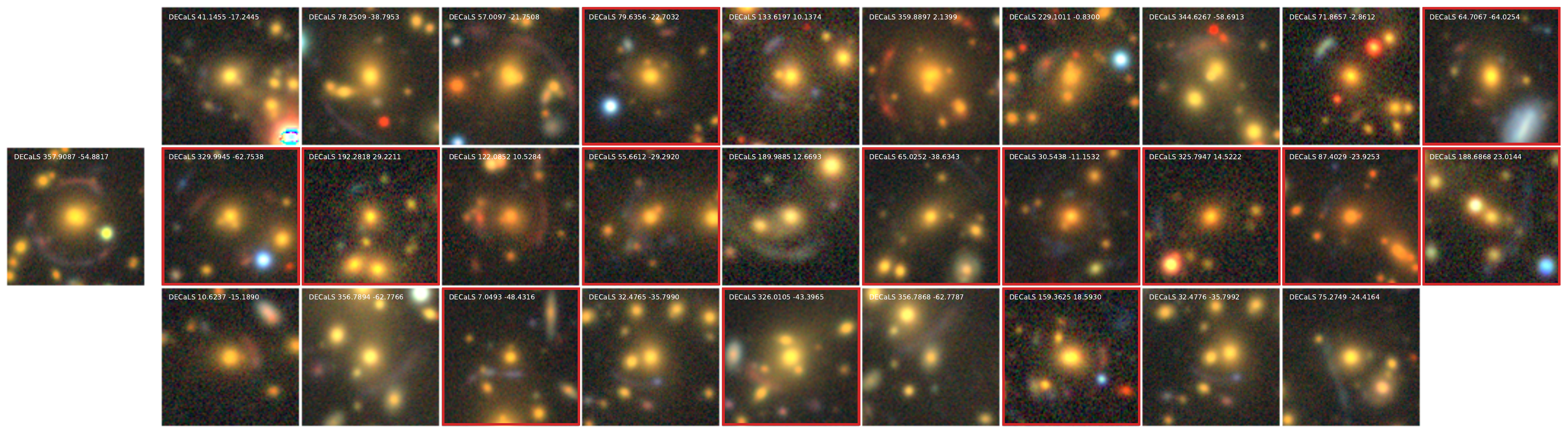}
    \includegraphics[width=1.0\textwidth]{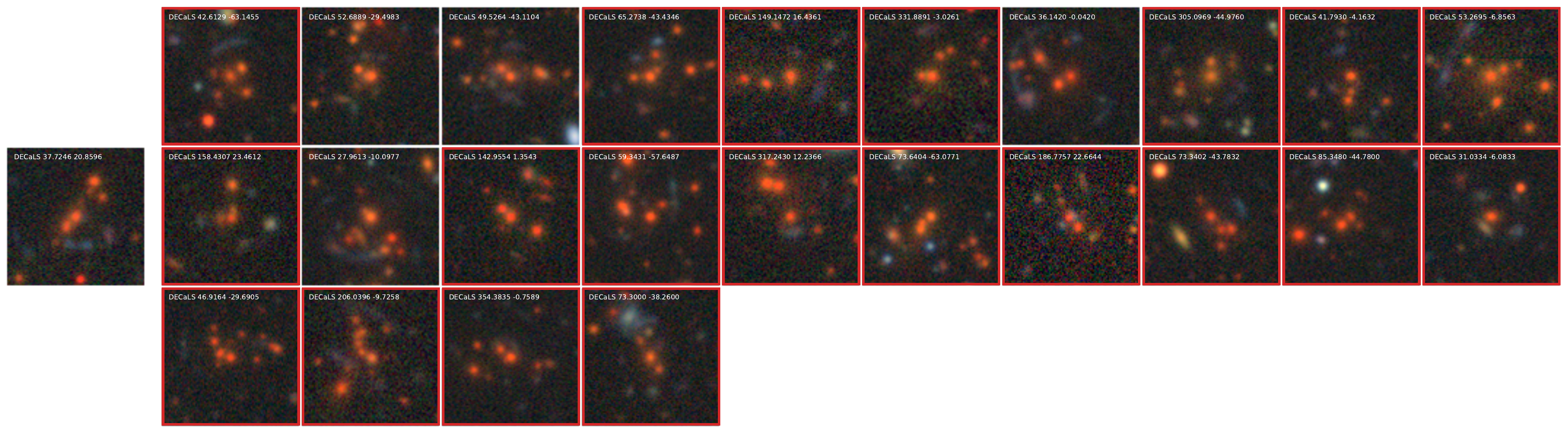}
    \includegraphics[width=1.0\textwidth]{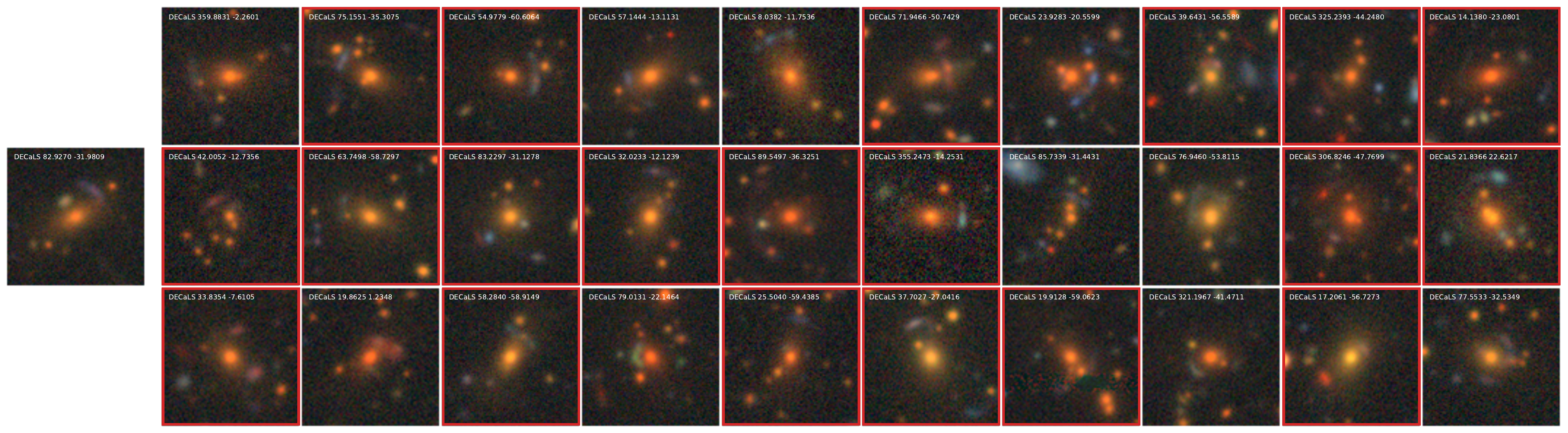}
    \caption{Similarity search for the discovery of new strong lens candidates, achieved without using a single label for training. We queried three images of previously discovered lenses (left), and on the right we display a sample of new strong lens candidates selected from the top 512 most similar images to each query, ordered by decreasing similarity score. Strong lens candidates with a red border are previously undiscovered. Using only three labelled data samples and $\sim$10 minutes of visual inspection we identified 83 strong lens candidates, 53 of them previously undiscovered.}
    \label{fig:similarity_search}
\end{figure*}

To assess our self-supervised model quality and search for new strong lens candidates in the DESI Legacy Surveys, we first employed a number of similarity searches using some of the known lenses identified in previous searches. We demonstrate this approach in Figure~\ref{fig:similarity_search}. For an illustrative example, we select three lenses from the training sample, each of different brightnesses and morphologies, and use the interactive similarity search tool we publicly provide to query for the 1000 most similar images. The first query image is a bright single central source with an extended and nearly complete lens arc, the second is a medium-brightness central cluster producing a small blue double-image arc, and the third is a single dim central galaxy producing an arc. For each query we visually examine the 512 most similar images, requiring only a few minutes, and flag any strong lens candidates. We find a total of 83 strong lens candidates given only three labelled queries - a significant number considering we have used only three examples and that our training sample of 1615 lenses constitutes all of the high-quality lenses in the south survey discovered to date. Of the 83 candidates found, to our knowledge, 53 of them have not been previously discovered. Compared to the probability of finding one strong lens in 10,000 images expected from a random sample of galaxies, self-supervised similarity search has immediately increased the lens frequency to $\sim$one in 20 (83:512$\times$3 vs 1:10,000), a factor of 500 speed-up. We found that some candidates were picked up by separate similarity searches of known lenses. 

The most similar objects returned by the search do not always share the same morphologies and colors as the labeled query, but feature ``similarity'' in a variety of characteristics. This result is most apparent in the first query presented of a bright central galaxy with an extended purple arc, where the most similar lenses returned display a wide variety of arc colors, arc sizes, and central source clustering but overall exhibit hard-to-quantify similar visual characteristics. As such, the 512 most similar objects we chose to visually inspect was an arbitrary cutoff, and additional lens candidates are likely to be uncovered by extending beyond this number.       

Similarity search can benefit supervised classification tasks in a number of ways, as a single label is sufficient to search for additional lens candiates, compared to the many hundreds or thousands required at a bare minimum for supervised training. Supervised classification is biased towards predicting the correct answer on the most common type of labels, and the predictive performance on rare occurrences suffers. Conversely, similarity search can be used to help rapidly identify additional members of less common examples, but may not guarantee the same degree of completeness as a full automated classification, as the new lens candidates by construction exhibit a high degree of visual similarity to the few labeled images provided. Thus, to maximize completeness, the techniques can be employed in tandem. Importantly, similarity search does not need to focus on finding additional lenses, and can instead be used to construct a more robust training sample of nonlenses. For example, ring galaxies, spiral galaxies, or galaxy mergers are commonly flagged as false positives by a number of previous lens finding studies due to their scarcity and similar appearance to gravitational lenses. While these objects are not as rare as strong lenses, when randomly sampling nonlenses they occur infrequently in smaller training sets, and the network can mistake them for lenses. With a similarity search, one can rapidly identify additional examples for these objects and include them in the training set to reduce the number of false positives during inference. 

With such results, the computational simplicity of a similarity search approach makes it a very attractive option for identifying new lenses, but it is worth highlighting limitations. Most notably, similarity search does not perform as well for images with distinct nonlens features offset from the central source that are unrelated to lensing, such as bright foreground stars or galaxies. For these cases the most similar images returned are those that also include similar foreground features, and the ``similarity focus'' has been drawn away from the lensing characteristics we are interested in. This is expected, as our self-supervised model is not given any explicit learning signal on what constitutes lensing features. One potential worka-round for this effect is to include such foreground effects as data augmentations in self-supervised pretraining --- for example, by randomly adding bright foreground objects to a given view of the original image --- which would make representations more invariant to these features. We leave this for future investigation.

We invite readers to explore our publicly available, web-based interactive similarity search tool and try to discover additional strong lenses beyond those presented here\footnote{\href{https://github.com/georgestein/ssl-legacysurvey}{github.com/georgestein/ssl-legacysurvey}}.

\subsection{Training Automated Lens Classification Models}
\label{sec:lens_classification_results}

\begin{figure}
    \centering
    \includegraphics[width=0.45\textwidth]{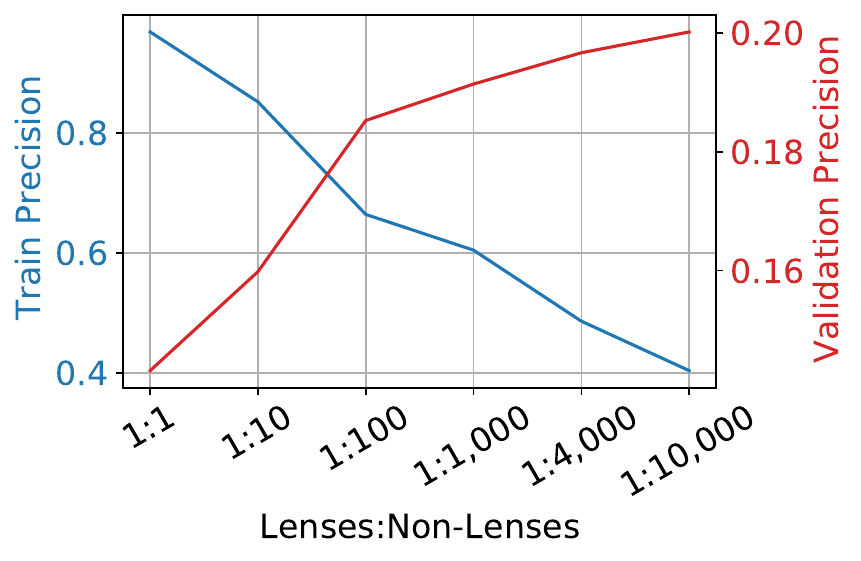}
    \caption{Linear classification precision as a function of the lens:nonlens ratio used for training. Validation is set at 1:10,000 to approximate the class imbalance expected in the full survey. As we increase the lens-to-nonlens ratio, we find lower performance on the training set but much higher performance on the validation data, demonstrating the advantage of training with class imbalances similar to what is expected in the true data.}
    \label{fig:trainingset}
\end{figure}

\begin{figure*}
    \centering
    \includegraphics[width=1.0\textwidth]{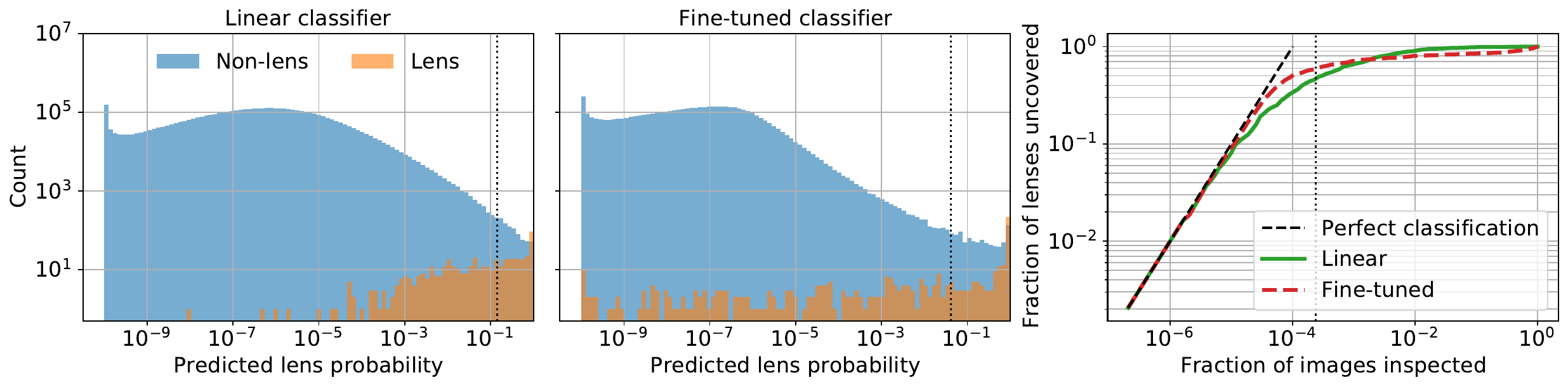}
    \caption{Predicted probability of image containing a lens over the validation set. The vertical dotted black lines designate the predictions over the portion of the survey that we expect to visually inspect in a few hours.}
    \label{fig:classification}
\end{figure*}

Given the large class imbalance between lensed and nonlensed images in the true data, it is critical to consider the balance of lenses and nonlenses in the training data for classification models. A sample with no class imbalance will include only a minimal number of negative examples (due to the small number of positive labels), which will not nearly cover the full distribution of galaxies in the dataset and will result in a larger number of false positives on the validation data. Alternatively, training using a significant class imbalance can result in poor convergence where the model does not predict any images as belonging to the positive class with any high degree of certainty. 

Using the 1,130 labelled lenses set aside for training, we sample six distinct training sets with increasing class imbalance, testing ratios of 1:1, 1:10, 1:100, 1:1000, 1:4000, and 1:10,000. When increasing the number of nonlenses, we sample additional galaxies while keeping those already included in smaller training sets (e.g., all representations from the 1:1 training set are also included in the 1:10 set). As shown in Figure~\ref{fig:trainingset}, we find a clear correlation between the lens-to-nonlenstraining ratio and the performance of the linear classifier on the validation set. When training with a much smaller degree of class imbalance than exists in the real data (1:1 through 1:100), we find that while the performance on the training set seems impressive, the performance on a realistic validation set suffers and the model predicts a significant number of false positives. As we increase the lens-to-nonlensratio, we find lower performance on the training set but much higher performance on the validation data, demonstrating the advantage of training with class imbalances similar to those expected in the true data. An increase of precision (Eq.~\ref{eq:precision}) from 0.14 (1:1) to 0.20 (1:4000) corresponds to a nearly 50\% decrease in the manual inspection time required to find a given number of strong lens candidates. Ratios of 1:4,000 and 1:10,000 give us our best results, and between the two, performance is comparable. Thus, for our remaining automated lens classification models we use the 1:4,000 ratio, as it is faster to train. 

With the training set finalized, linear classification proceeds in a straightforward manner. For fine-tuning we need to additionally determine the optimal lens oversampling factor. As explained in Section \ref{sec:finetuning_methods}, oversampling is required to ensure that each minibatch of 512 samples includes both positive and negative instances. We varied the oversampling factor to ensure that each minibatch, on average, contains (1\%, 2.5\%, 5\%, 10\%) of its samples from the positive class (lens), and found that 5\% achieved best performance on the validation set. This corresponds to an oversampling factor of 40, such that during an epoch each lens is seen 40 times (with different augmentations applied for each sample), while nonlenses are only seen once per epoch. Linear classification required a few minutes on a CPU, while fine-tuning used 36 hours on eight GPUs.

We show the predicted lens probability over the validation set of the linear classification and fine-tuned models in Figure~\ref{fig:classification}. The results of linear classification and fine-tuning are quite similar, with roughly half the true lenses in the validation set receiving probabilities above the $N=1139$ cutoff (dotted vertical black lines) determined by the number of images that we have time to visually inspect. For both models we find a clear correlation between the visual quality of the lens candidates and the predicted lens probability. Images with clear features of strong gravitational lensing, such as distinctive arcs, are almost always given a high predicted lens probability, while lens candidates with faint features are given low predicted values. We also find a great agreement between the predictions of the linear classifier and fine-tuned models for individual image/representation pairs -- both models identify similar lenses. 

In the right hand panel of the Figure~\ref{fig:classification} we estimate the number of images that one needs to visually inspect in order to find a given fraction of lenses, assuming that our validation set is representative of the true distribution and there is one lens in every 10$^4$ galaxies. The initial validation set will contain a number of undiscovered lenses that introduce a certain degree of mislabelling, and we later confirm that a significant number of the ``nonlenses'' with a high predicted lens probability are in fact previously undiscovered strong gravitational lenses. From the figure we see that the linear classifier and fine-tuned network perform nearly perfectly on the top 10\% of lenses, and identify 50\% and 60\% of the lenses above the manual inspection cutoff, respectively. As we extrapolate to the full DECaLS \zmag $< 20$ sample, the fine-tuned network slightly outperforms the linear classifier if inspecting the top 10$^4$ predictions (0.02\% of the total), but the linear classifier in fact performs better than the fine-tuned if inspecting more than the top 10$^5$ predictions (0.2\% of the total), due to slight overfitting of the fine-tuned model. These estimates do not account for any undiscovered lens configurations that do not have similar examples in the labeled data set.

\subsection{Searching for new strong lens candidates}

After the linear classification and fine-tuned models were trained using the DECaLS \zmag $< 20$ training sample we applied them to the remainder of the DECaLS \zmag $<$ 20 sources, and the full north dataset. Although the \zmag $>$ 20 sample was dimmer than any data seen during training we found that both models extrapolated well, and resulted in accurate predictions upon visual inspection. The north survey was observed with different instruments and bandpasses and had higher noise levels than the south, thus introducing a train-test discrepancy to such experiments. Nevertheless, as we will demonstrate in the following section, models trained only on south images and applied to the north survey also resulted in the detection of a number of new strong lens candidates. 

Using the linear classification and fine-tuned models a combination of authors J.B., T.M. and G.S. inspected:
\begin{itemize}
    \item $\sim$25 similarity searches using lenses selected from the training set,
	\item the top $\sim$7,500 predictions over the 43 million DR9 south galaxies with \zmag $<$ 20,
    \item the top $\sim$3,000 predictions over 20 million DR9 south galaxies with \zmag $>$ 20,
    \item the top $\sim$5,000 predictions over DR9 north.
\end{itemize}
After an initial test, where all three authors examined an identical set of predictions to ensure consistency between visual classifications, the samples outlined above were split into chunks, and there was little overlap of galaxies inspected by the different authors. The top linear classification and fine-tuned predictions had a significant overlap, and the linear classification model received the most attention. Fine-tuned predictions were inspected after the linear classification to identify any missed lenses in the first pass, or any that did not make the linear classification cutoff. We discarded any overlapping lens candidates that resulted from nearby galaxies of the same cluster being independently identified as lenses by different methods. 

Extrapolating from Figure~\ref{fig:classification}, we expect that in the south \zmag $<$ 20 sample we have identified 50-60\% of the lenses, and 30-40\% remain. Approximately 20\% of the remaining lenses are likely to be in the top 100,000 predictions, which is an achievable number to manually inspect. This is supported by the fact that we continued to find a number of high-quality lenses after the 7000th highest prediction. The lens candidates remaining to be discovered will likely not be configurations with extremely distinct extended arcs, as we found that those were generally predicted with the highest probabilities in our validation sample. The number of remaining lenses in the south \zmag~$>$~20 or north samples is more difficult to estimate, but we expect is greater than the fractions quoted above.

We separated our new strong lens candidates into two groups based on their quality. During the lens search outlined above, the authors assigned each lens candidate either a tentative A or B grade, which a single author, G.S., refined into the final lens classifications to achieve consistency. Grade A lens candidates are highly likely to be strong gravitational lenses as they exhibit distinct lens features such as large extended arcs or clear multiple images. Grade B galaxies exhibit fainter features than grade A, and are more ambigious with common false positives such as ring galaxies. Nevertheless, grade B lenses still display likely features of gravitational lensing. To include galaxies as either grade A or B strong lens candidates we erred on the side of caution, and attempted to exclude candidates that displayed characteristics of gravitational lensing but were difficult to differentiate from projection or noise effects. As there is no distinct criteria for lens classification the small number of nonexpert galaxy inspectors likely introduced biases to the type of lenses identified in this work, and possibly passed over certain groups of objects. For this reason, we make the network predictions and similarity search tool public and available for anyone to examine.

\subsection{Catalogue of new strong lens candidates}

\begin{table}
\centering
\begin{tabular}{c|l|l}
\textbf{Grade} & \textbf{Survey} & \textbf{New Lens Candidates} \\ 
\hline
& South \zmag $<$ 20 & 316 \\
A & South \zmag $>$ 20 & 47 \\
& North & 41 \\
& All & 404 \\
\hline
& South \zmag $<$ 20& 583 \\
B & South \zmag $>$ 20  & 131 \\
& North & 74 \\
& All & 788 \\
\hline
Total & & 1192 \\
\end{tabular}
\caption{New strong gravitational lens candidates split by survey region and lens grade.}
\label{tab:lenses}
\end{table}

\begin{figure*}
    \centering
    \includegraphics[width=0.98\textwidth]{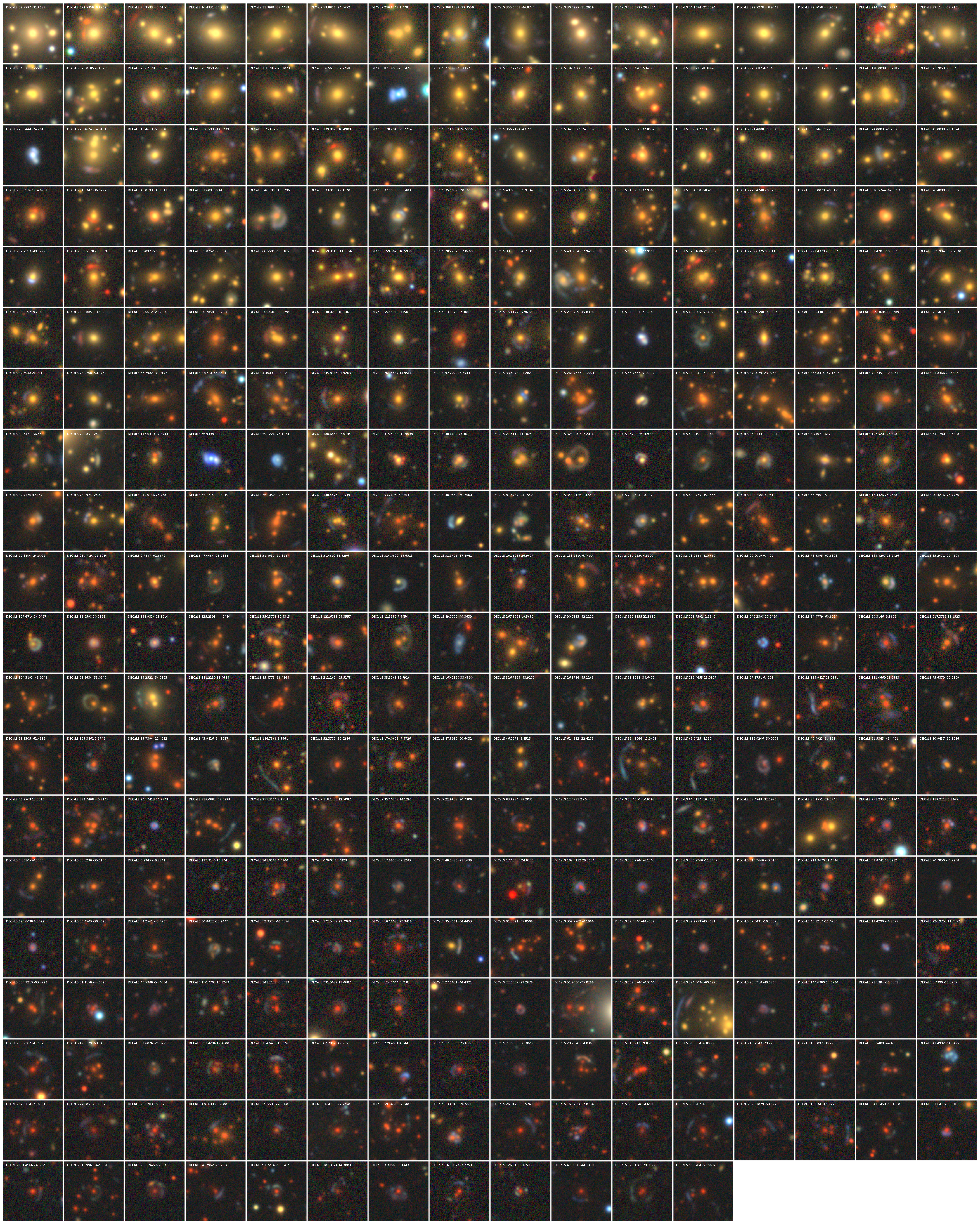}
    \caption{South survey grade A lens candidates with \zmag $<$ 20.}
    \label{fig:a_south}
\end{figure*}

\begin{figure*}
    \centering
    \includegraphics[width=0.98\textwidth]{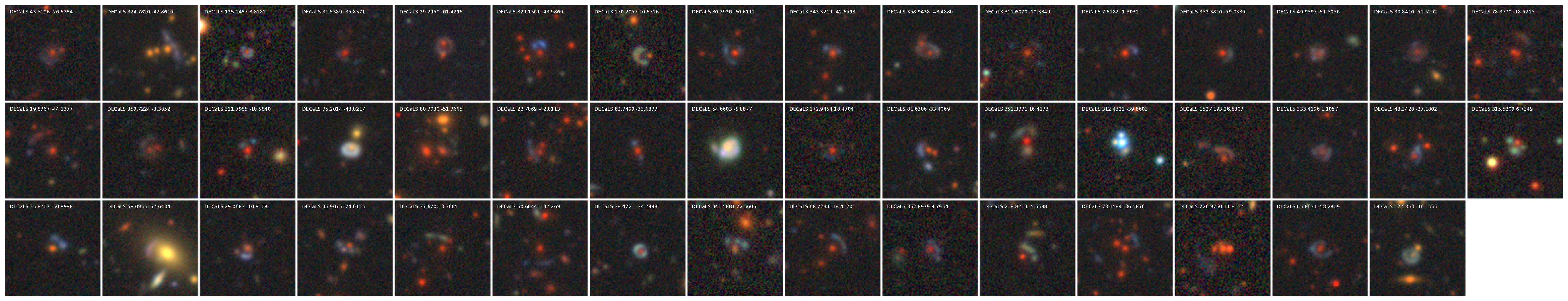}
    \caption{South survey grade A lens candidates with z-band magnitude $>$ 20.}
    \label{fig:a_southmag20-21}
\end{figure*}
\begin{figure*}
    \centering
    \includegraphics[width=0.98\textwidth]{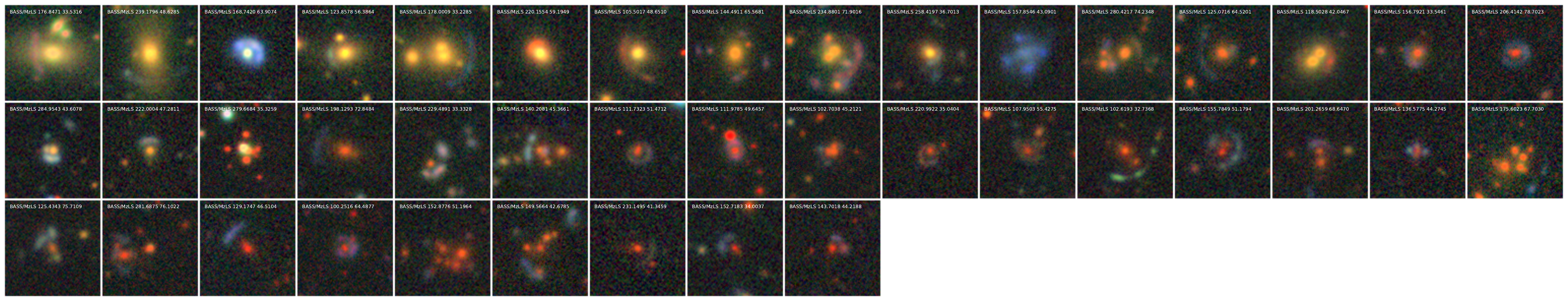}
    \caption{North survey grade A lens candidates.}
    \label{fig:a_north}
\end{figure*}

We compiled the lens candidates discovered through similarity search, linear classification, and fine-tuning into a single catalogue consisting a total of 1192 new strong lens candidates. Table ~\ref{tab:lenses} outlines the number of candidates split by lens grade and survey region. The majority of new lens candidates are from the south survey and have \zmag $<$ 20. This is expected as the south survey, specifically the DES region, has a lower noise level than the north, and was also the subset of the data used to train the models. We display the 404 grade A lens candidates in Figures~\ref{fig:a_south} through~\ref{fig:a_north}, split by survey region, and show the B candidates in Appendix~\ref{app:additional_lenses}.

The most common grade A lens configurations are arcs separated by at least a few arcseconds from central source(s), more extended arcs separated by a smaller angular distance from a central source, and duplicate background galaxies. We also find numerous Einstein crosses and other rare lens configurations. The majority of central galaxies have a red or orange appearance, but we also find a number of lensing features around blue central galaxies. The diversity of lens candidates discovered here demonstrates the power of a purely data-driven self-supervised approach capable of learning from every single galaxy in the survey. Our method did not require any cuts on the galaxy or lens sample used; thus, we are able to easily identify lens candidates covering the full variety of appearances. In contrast, the majority of previous studies subsample the dataset to include only luminous red galaxies (LRGs) by introducing hand-selected color and magnitude cuts, which, given their redshift distribution and masses are the most likely type of galaxy to lens background sources \citep{1984ApJ...284....1T}. Targeting only a subsample of the full range of gravitational lenses in the universe will only allow for classification of the given specific lens type, and not generalize outside of the limited distribution of data used for training.      

\begin{figure*}
    \centering
    \includegraphics[width=1.0\textwidth]{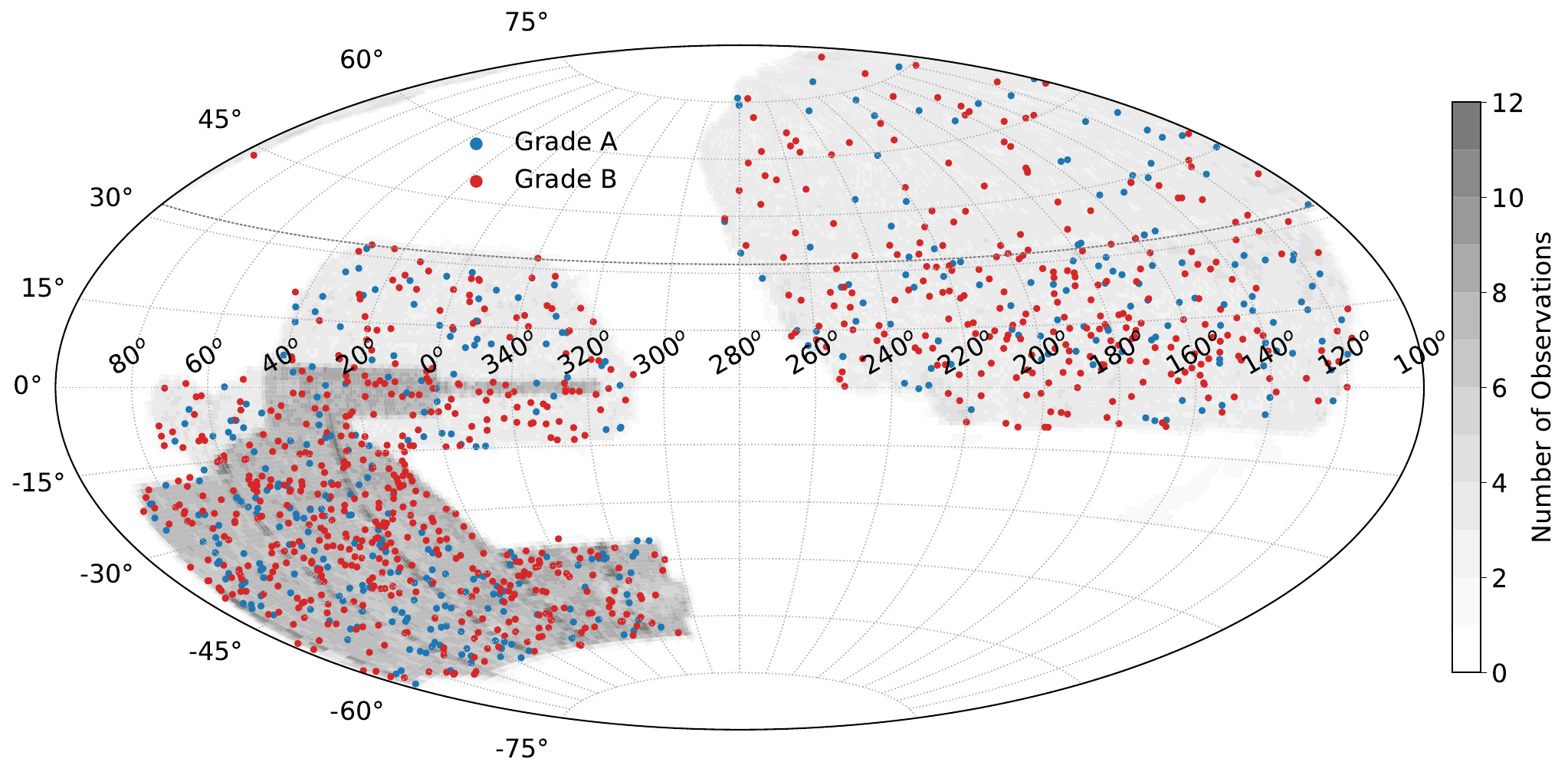}
    \caption{Right ascension and declination of the new strong lens candidates, with the background shaded by the average number of observations across the three photometric bands in each square degree region. The gray line at $\delta=32.375$ denotes the split between the north and south surveys.}
    \label{fig:sky_plots}
\end{figure*}

\begin{figure*}
    \centering
    \includegraphics[width=1.0\textwidth]{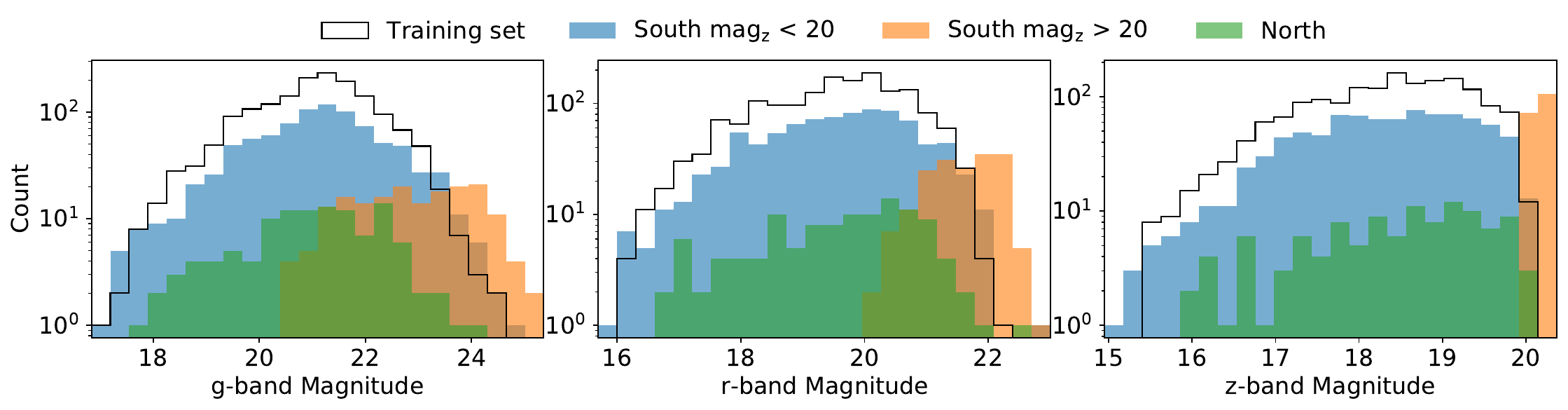}
    \begin{tabular}{p{0.33\textwidth} p{0.345\textwidth}}
      \vspace{0pt} \includegraphics[width=0.33\textwidth]{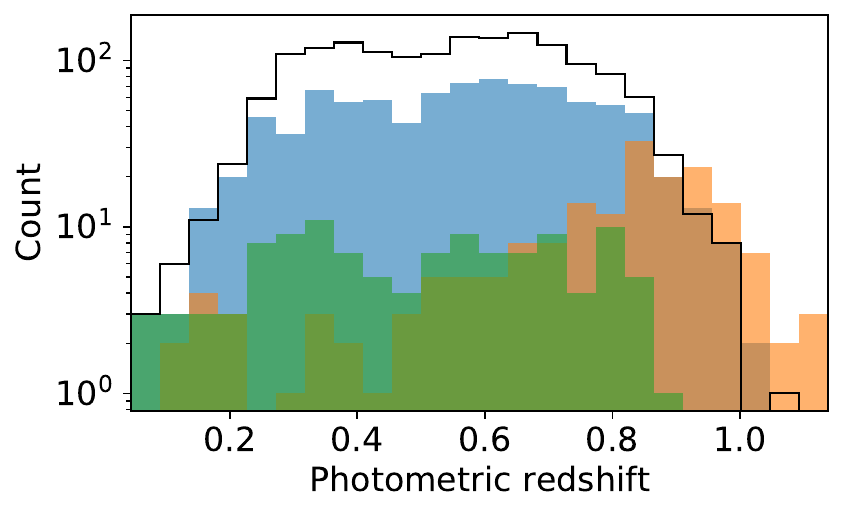} &
      \vspace{0pt} \includegraphics[width=0.345\textwidth]{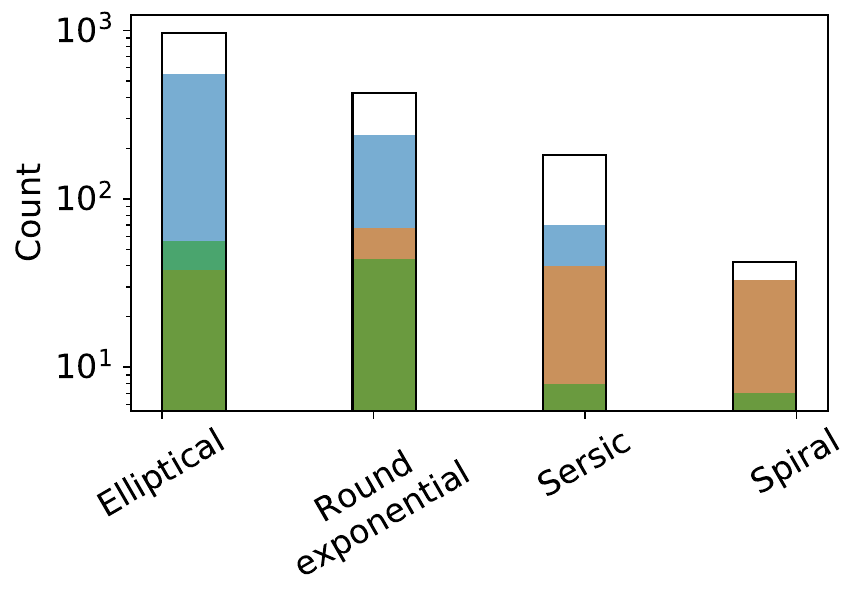}
    \end{tabular}
    \caption{New strong lenses by central source magnitudes, photometric redshifts from \citet{rongpu}, and the central lens morphology from the Tractor fits.}
    \label{fig:histograms}
\end{figure*}

Figure~\ref{fig:sky_plots} illustrates the distribution of the new strong lens candidates on the sky in right ascension and declination, with the background shaded by the number of observations used to construct DR9 images. We find that the DES region (depicted clearly in the bottom left as the dark gray region with declination $<$ 5\degree) contains the highest density of new lenses. This region was imaged to the greatest depth and thus contains the lowest level of noise, which means that lenses should be more visually distinct and easy to identify. This low noise level also means that the DES region has been the focus of the majority of recent previous lens finding works, and thus many of the lenses have already been found. Nevertheless, in this work we find a large number of lens candidates in this region that were not identified in any other search. The most significant number of lenses discovered in DESI Legacy survey data was achieved by \citep{HuangII}, who used the previous data release, DR8, which had incomplete coverage below $\delta \approx -32\degree$. Although in Figure~\ref{fig:sky_plot_data} we see a much greater density of training samples above this line than below, we do not find the same with our new lenses. This means we are not simply finding lenses in regions of the sky that have not been searched, but that we are identifying lenses missed by previous works. It is also clear that the north region is the least densely populated, both the due to increased level of noise, and the fact that our models were trained on south images yet applied to the north even though north images used slightly different photometric bands. We find no distinct correlations between the sky distribution and the lens grade, photometric redshift, or z-band magnitude. 

We summarize the distribution of magnitudes, photometric redshifts, and the source morphology of the lenses in Figure~\ref{fig:histograms}, split by survey region. We find that the newly discovered lenses closely resemble the training set, with the exception of the \zmag~$>$~20 south lens candidates which are by definition dimmer than the set used for training. The majority of newly discovered lens candidates are elliptical galaxies (deVaucouleurs =``DEV''), and number are around spiral galaxies (exponential=``EXP''). Round exponential galaxies with a variable radius (``REX'') and Sersic profiles (``SER'') are more rare, but we still find a number with these morphological types.

This new catalogue of 1192 strong gravitational lens candidates is summarized in data tables in Appendix~\ref{app:additional_lenses} 

\section{Conclusion}
\label{sec:conclusions}

In this work, we demonstrated the first direct application of self-supervised machine learning for finding rare objects in large astronomical datasets, finding 1192 new strong gravitational lens candidates in the DESI Legacy Imaging Surveys. After training the self-supervised model on a large body of unlabeled data, we presented three different methods for the discovery of strong lenses -- similarity search, linear classification, and fine-tuning -- each with their own distinct advantages over a standard supervised classification setup. The similarity search tool provides an avenue to rapidly identify additional lenses given just a single labeled sample (see Figure~\ref{fig:similarity_search}), which is orders of magnitude less labeled data than would be required to train even the simplest supervised network. Additionally, as the self-supervised model was only trained to distinguish the natural distribution of morphologies and features within the survey sample, it is not limited to  identification of a  specific class of object but can be used to facilitate the discovery of any object(s) that exist in the dataset.

Linear classification and fine-tuning the self-supervised network both proved extremely useful for the automated classification of new strong lens candidates. Although supervised fine-tuning slightly improved the classification results over the fraction of the dataset we visually inspected, it is important to consider and contrast the computational resources, the human time investment, and the machine learning expertise required to perform the two types of automated searches, and what this means for survey science. Supervised training of the CNN in a distributed GPU setup provides a number of barriers to rapid scientific discovery; it requires a nonnegligible machine learning background, access to a machine capable of storing the full image data on disk (2.5 TB for the training and validation set images and 19 TB for the full DESI Legacy Survey sample used in this work), and access to top-of-the-line GPUs for days just for the final network training (which does not including the time required for hyperparameter searches). Linear classification provides a compelling alternative; it requires only CPU access and minimal machine learning or computational background, and the distilled representations can fit on any modest machine (76 GB for the training and validation sets and 579 GB for the full survey sample). 

Using a combination of the three methods, we conducted a fast visual inspection campaign of $\sim$18,000 DESI Legacy Survey images, from which we identified 1192 new strong lens candidates. This is a significant number when compared to previous supervised CNN studies, despite comparatively little time dedicated to visual identification. For example, using more restrictive data cuts of the DESI Legacy Survey DR8 that prefilter galaxies not likely to be lenses, \citealt{HuangI} found 335 new candidates in $\sim$50,000 inspections, and \citealt{HuangII} found 948 new candidates in 38,679 inspections of their main ResNet model. We emphasize that direct comparisons between our study and such previous works are difficult due to two competing effects, and as such, these numbers are stated here only as a guideline. On one hand, previous lens searches that utilized previously discovered lenses as their training sample (rather than a simulated sample \citep[e.g.]{JacobsI}, in which lenses are adequately numerous but may lack full realism) generally had a smaller number of labelled lens candidates to train their models, which limits their classification performance. But, on the other hand, previous searches had a greater number of undiscovered high-quality lenses in the unmined data, and the number of remaining lenses distinguishable at the resolution of the DESI Legacy Surveys has diminished with each successive lens search \cite{HuangI, HuangII}.

We expect that a nontrivial number of lenses in our catalog, on the order of 30-40\% based on the validation set experiment, remain undiscovered.  However, they can be discovered by subsequent similarity searches or inspections of network predictions that fell outside of our limited labelling campaign. For this reason, we are releasing the similarity search tool and the network predictions to facilitate the identification of additional strong lenses through crowd-sourcing\footnote{Predictions and the location of the similarity search tool can be found at \href{https://github.com/georgestein/ssl-legacysurvey}{github.com/georgestein/ssl-legacysurvey}}.

While the lens search was highly successful, perhaps this is the most important result of this work; the self-supervised model was trained in a generalized task-agnostic way requiring no labelled samples, yet the resulting representations enabled rapid and easy discovery of specific and rare objects. We emphasize that this greatly reduces the barrier to entry for working with modern survey datasets and creates the potential to open up a number of collaborative avenues that were previously unavailable. Rather than each team working alone to perform classification or regression tasks on survey datasets -- applying for time on large GPU computing systems, downloading massive datasets, learning to train models in parallel, etc. -- the initial model training can be undertaken separately and easily shared to allow for rapid scientific investigations that are not hindered by computational resource requirements.
 
\section{Acknowledgements}
We would like to thank Rongpu Zhou for his significant help in using the DESI Legacy Survey data, Dustin Lang for providing access to the image-cutout service at NERSC, and Md Abul Hayat and Mustafa Mustafa for their pioneering efforts on self-supervised learning for sky surveys. We would also to thank Solene Chabanier for help with lens labelling, and Xiaosheng Huang, Andi Gu, Christopher Storfer, and Saurav Banka for discussions on lens finding.

This research used resources of the National Energy Research Scientific Computing Center (NERSC), a U.S. Department of Energy Office of Science User Facility located at Lawrence Berkeley National Laboratory, operated under Contract No. DE-AC02-05CH11231. This research also used resources of the Argonne Leadership Computing Facility, which is a DOE Office of Science User Facility supported under Contract DE-AC02-06CH11357. G.S., J.B., T.M., and Z.L.~were partially supported by the DOE's Office of Advanced Scientific Computing Research and Office of High Energy Physics through the Scientific Discovery through Advanced Computing (SciDAC) program.

The Legacy Surveys consist of three individual and complementary projects: the Dark Energy Camera Legacy Survey (DECaLS; Proposal ID \#2014B-0404; PIs: David Schlegel and Arjun Dey), the Beijing-Arizona Sky Survey (BASS; NOAO Prop. ID \#2015A-0801; PIs: Zhou Xu and Xiaohui Fan), and the Mayall z-band Legacy Survey (MzLS; Prop. ID \#2016A-0453; PI: Arjun Dey). DECaLS, BASS and MzLS together include data obtained, respectively, at the Blanco telescope, Cerro Tololo Inter-American Observatory, NSF’s NOIRLab; the Bok telescope, Steward Observatory, University of Arizona; and the Mayall telescope, Kitt Peak National Observatory, NOIRLab. The Legacy Surveys project is honored to be permitted to conduct astronomical research on Iolkam Du’ag (Kitt Peak), a mountain with particular significance to the Tohono O’odham Nation.

NOIRLab is operated by the Association of Universities for Research in Astronomy (AURA) under a cooperative agreement with the National Science Foundation.

This project used data obtained with the Dark Energy Camera (DECam), which was constructed by the Dark Energy Survey (DES) collaboration. Funding for the DES Projects has been provided by the U.S. Department of Energy, the U.S. National Science Foundation, the Ministry of Science and Education of Spain, the Science and Technology Facilities Council of the United Kingdom, the Higher Education Funding Council for England, the National Center for Supercomputing Applications at the University of Illinois at Urbana-Champaign, the Kavli Institute of Cosmological Physics at the University of Chicago, Center for Cosmology and Astro-Particle Physics at the Ohio State University, the Mitchell Institute for Fundamental Physics and Astronomy at Texas A\&M University, Financiadora de Estudos e Projetos, Fundacao Carlos Chagas Filho de Amparo, Financiadora de Estudos e Projetos, Fundacao Carlos Chagas Filho de Amparo a Pesquisa do Estado do Rio de Janeiro, Conselho Nacional de Desenvolvimento Cientifico e Tecnologico and the Ministerio da Ciencia, Tecnologia e Inovacao, the Deutsche Forschungsgemeinschaft and the Collaborating Institutions in the Dark Energy Survey. The Collaborating Institutions are Argonne National Laboratory, the University of California at Santa Cruz, the University of Cambridge, Centro de Investigaciones Energeticas, Medioambientales y Tecnologicas-Madrid, the University of Chicago, University College London, the DES-Brazil Consortium, the University of Edinburgh, the Eidgenossische Technische Hochschule (ETH) Zurich, Fermi National Accelerator Laboratory, the University of Illinois at Urbana-Champaign, the Institut de Ciencies de l’Espai (IEEC/CSIC), the Institut de Fisica d’Altes Energies, Lawrence Berkeley National Laboratory, the Ludwig Maximilians Universitat Munchen and the associated Excellence Cluster Universe, the University of Michigan, NSF’s NOIRLab, the University of Nottingham, the Ohio State University, the University of Pennsylvania, the University of Portsmouth, SLAC National Accelerator Laboratory, Stanford University, the University of Sussex, and Texas A\&M University.

BASS is a key project of the Telescope Access Program (TAP), which has been funded by the National Astronomical Observatories of China, the Chinese Academy of Sciences (the Strategic Priority Research Program “The Emergence of Cosmological Structures” Grant \# XDB09000000), and the Special Fund for Astronomy from the Ministry of Finance. The BASS is also supported by the External Cooperation Program of Chinese Academy of Sciences (Grant \# 114A11KYSB20160057), and Chinese National Natural Science Foundation (Grant \# 11433005).

The Legacy Survey team makes use of data products from the Near-Earth Object Wide-field Infrared Survey Explorer (NEOWISE), which is a project of the Jet Propulsion Laboratory/California Institute of Technology. NEOWISE is funded by the National Aeronautics and Space Administration.

The Legacy Surveys imaging of the DESI footprint is supported by the Director, Office of Science, Office of High Energy Physics of the U.S. Department of Energy under Contract No. DE-AC02-05CH1123, by the National Energy Research Scientific Computing Center, a DOE Office of Science User Facility under the same contract; and by the U.S. National Science Foundation, Division of Astronomical Sciences under Contract No. AST-0950945 to NOAO.

The Photometric Redshifts for the Legacy Surveys (PRLS) catalog used in this paper was produced thanks to funding from the U.S. Department of Energy Office of Science, Office of High Energy Physics via grant DE-SC0007914.

\appendix
\section{Strong lens catalogue and additional lens candidates}
\label{app:additional_lenses}

\begin{figure*}
    \centering
    \includegraphics[width=0.98\textwidth]{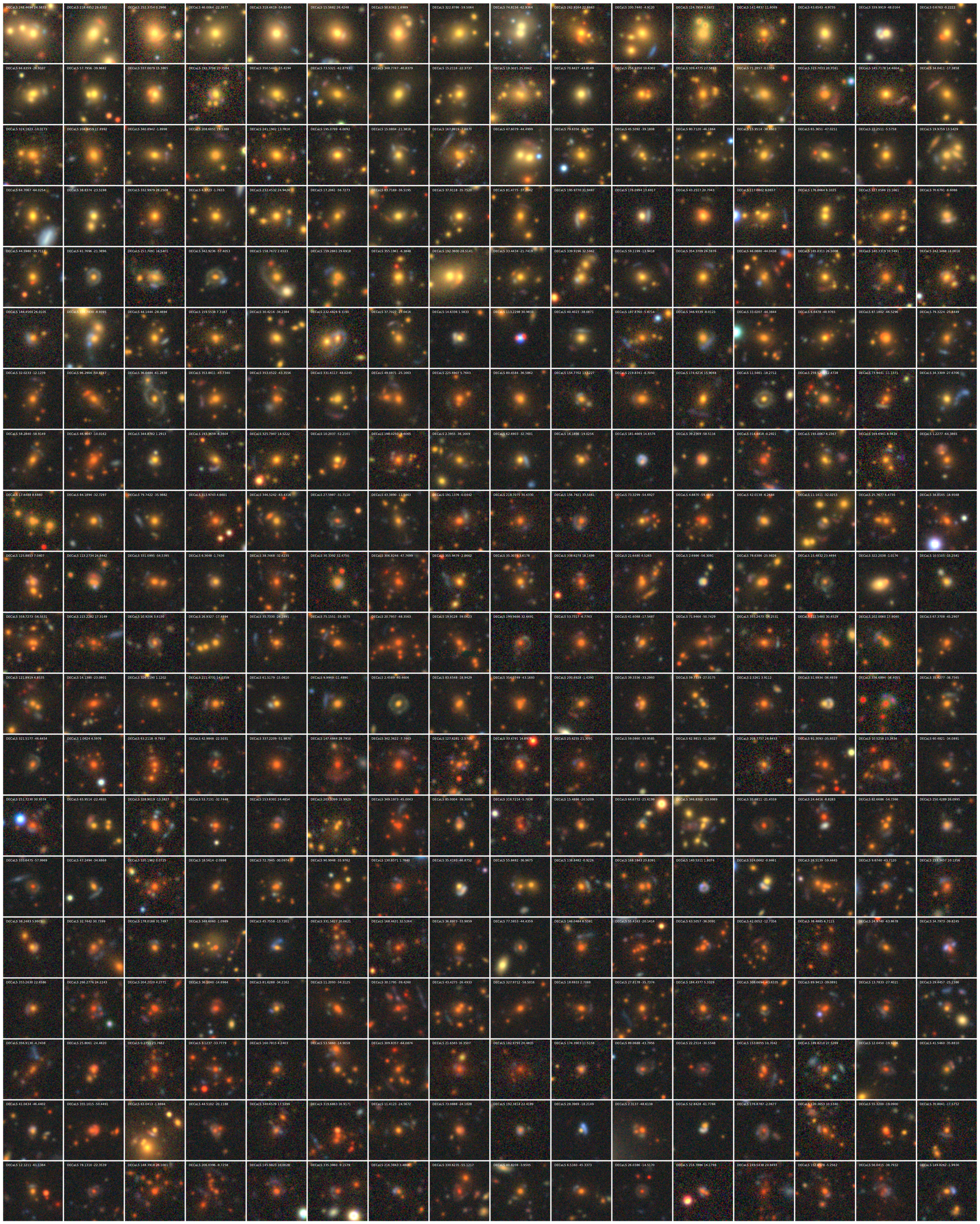}
    \caption{South survey grade B lens candidates with z-band magnitude $<$ 20.}
    \label{fig:b_south_1}
\end{figure*}

\begin{figure*}
    \centering
    \includegraphics[width=0.98\textwidth]{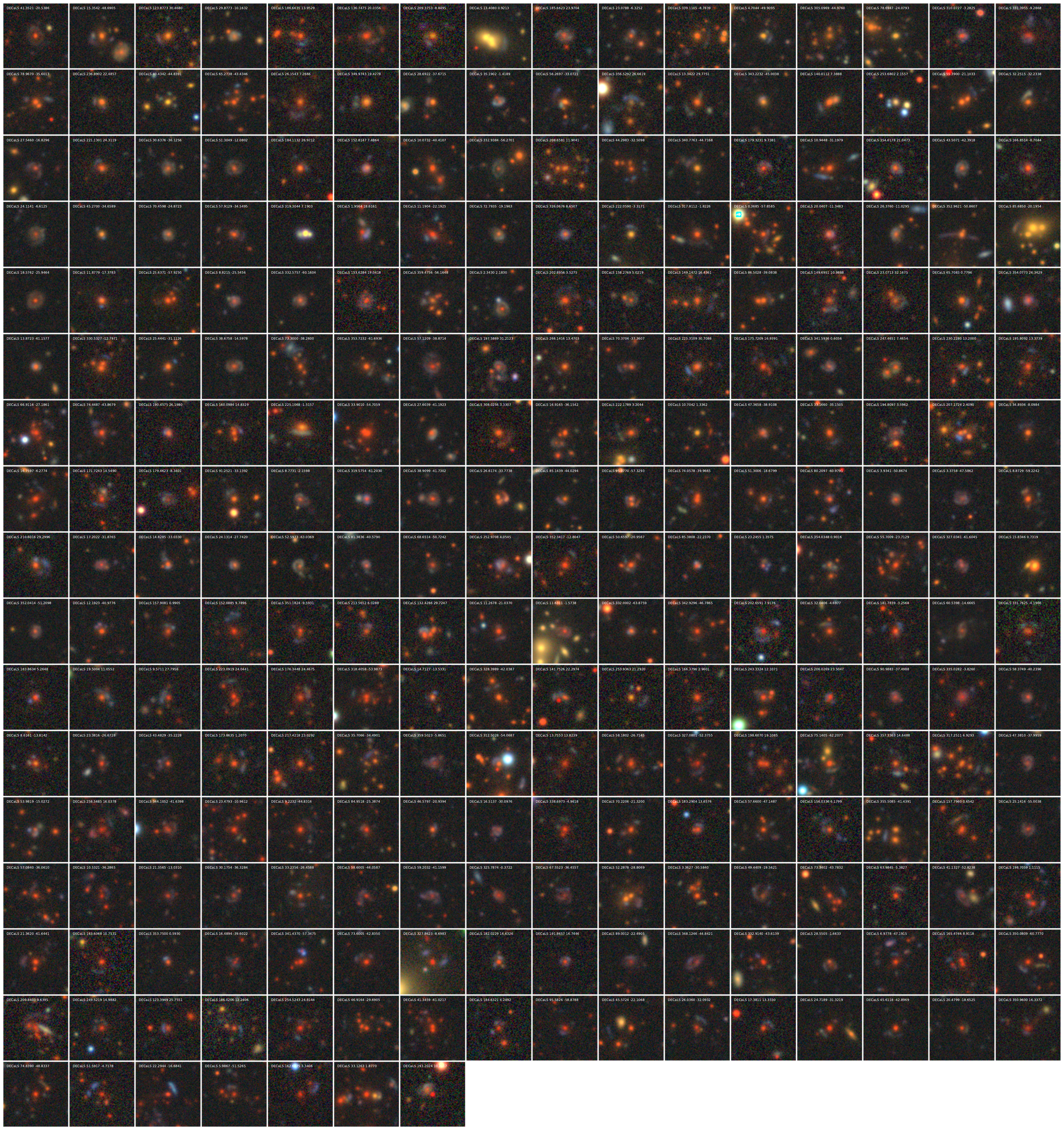}
    \caption{Figure~\ref{fig:b_south_1} continued: south survey grade B lens candidates with z-band magnitude $<$ 20.}
    \label{fig:b_south_2}
\end{figure*}

\begin{figure*}
    \centering
    \includegraphics[width=0.98\textwidth]{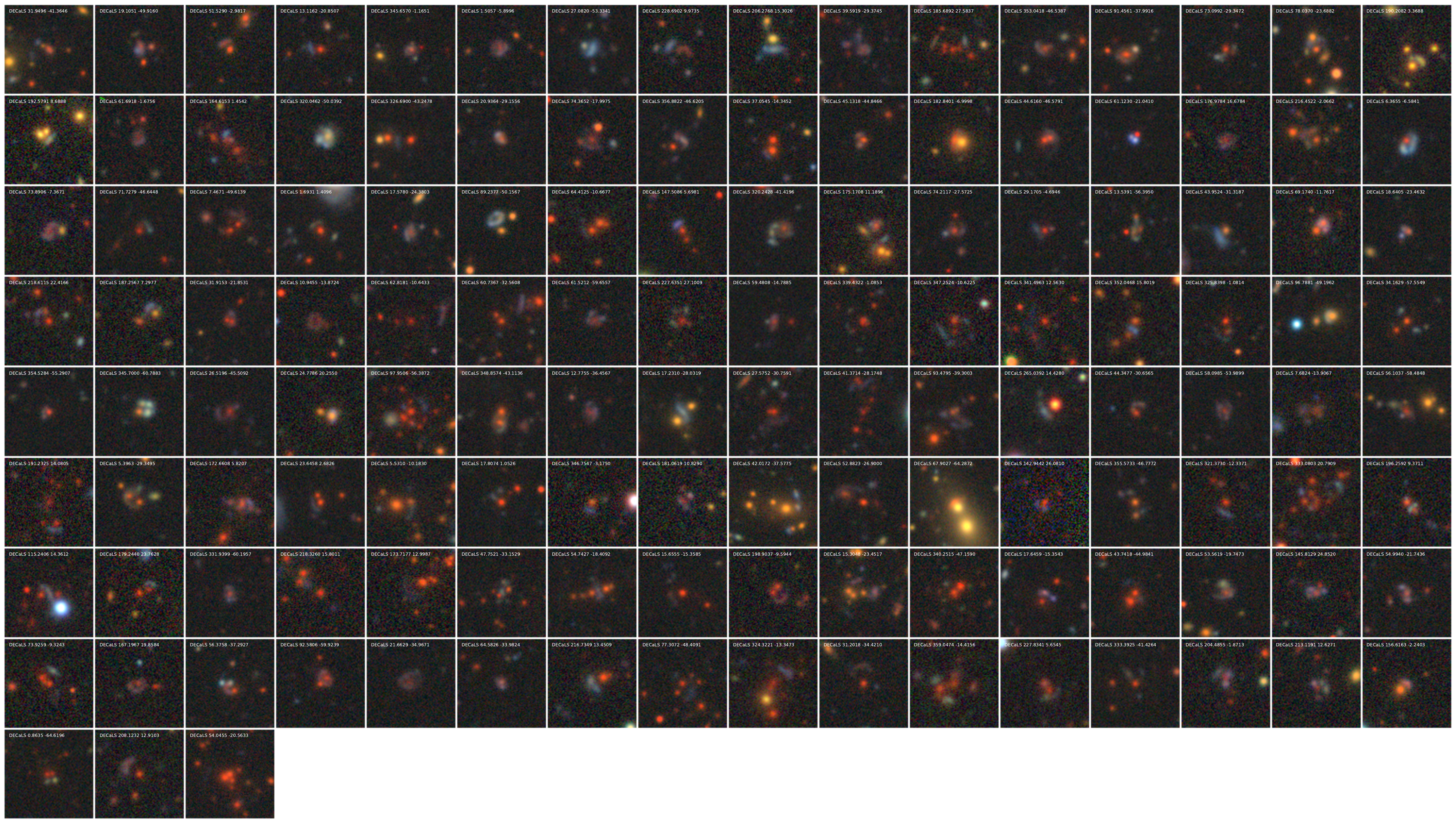}
    \caption{South survey grade B lens candidates with z-band magnitude $>$ 20.}
    \label{fig:b_southmag20-21}
\end{figure*}

\begin{figure*}
    \centering
    \includegraphics[width=0.98\textwidth]{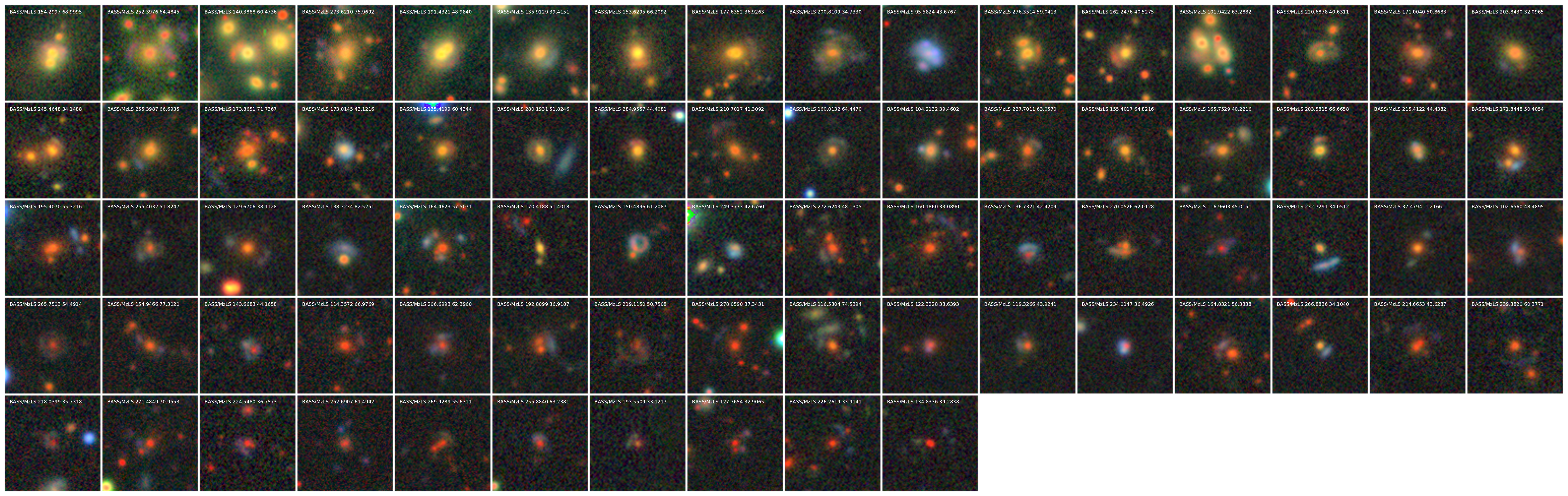}
    \caption{North survey grade B lens candidates.}
    \label{fig:b_north}
\end{figure*}

We provide the 1192 new lens candidates  identified in this work in data tables at \href{https://github.com/georgestein/ssl-legacysurvey}{github.com/georgestein/ssl-legacysurvey}, and display the grade B lenses in Figures~\ref{fig:b_south_1} through ~\ref{fig:b_north}. Due to the fainter lens features of Grade B lenses compared to grade A, it is likely that the grade B catalog contains a larger fraction of false positives. For this reason we recommend that follow-up observations and investigations start with the grade A catalogue before moving onto the B. During writing this article, after the conclusion of our lens search,   \citet{rojas2021strong} performed an additional supervised CNN search on the DES region, finding 186 new lens candidates after examining their top 76,582 network predictions. We have not checked for overlap with our new lens candidates.


\section{Data Augmentations}
\label{app:augmentations}
\begin{figure}
    \centering
    \includegraphics[width=0.5\textwidth]{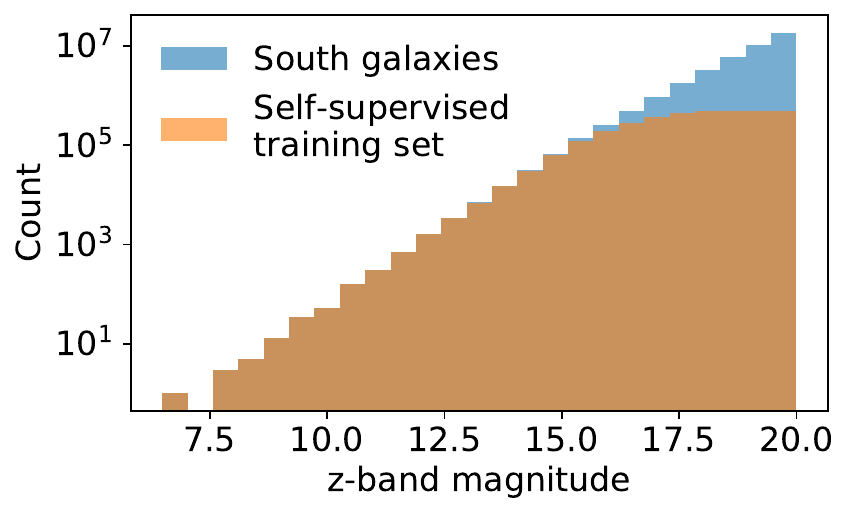}
    \caption{z-band magnitude of the 3,500,000 galaxy sample used for training the self-supervised model compared to the full distribution. Although the brightest and most nearby galaxies will not constitute strong lenses we do not use any minimum magnitude cut to remove them from the self-supervised training set - the goal at this stage is only to learn task-agnostic galaxy representations from a diverse set of galaxy images.}
    \label{fig:ssl_training_set}
\end{figure}

To design the data augmentations we fit to measured distributions of the Gaussian noise level, point spread smoothing, and extinction in images from the south survey sample with \zmag $<$ 20, as this was the sample used to train the self-supervised model. Distributions were fit separately in g, r, and z bands, and the augmentations were considered to be uncorrelated across bands. This is a reasonable approximation given that each image is compiled from a number of different observations taken at different times.

To determine the level of Gaussian noise in DR9 images we used the noise equivalent area (NEA) calculated using the blob-masked version outlined at \href{https://www.legacysurvey.org/dr9/nea/}{legacysurvey.org/dr9/nea/},
\begin{equation}
    \frac{1}{\sigma^2_{pix}} = 4\pi\ \mathrm{psfdepth} \left[ \frac{\mathrm{psfsize}}{2.3548}\right]^2,
\end{equation}
where psfsize is in units of pixels (0.262 arcsecond), and $\sigma_{pix}$ is estimated Gaussian noise level in each pixel. The standard deviation of the PSF was calculated by 
\begin{equation}
    \sigma_{\mathrm{PSF}} = \frac{\mathrm{psfsize}}{2.3548},
\end{equation}
where psfsize is again in units of pixels, and the extinction was given in the sweep catalogues. Examining the histogram of the values measured from the 43 million galaxies we found that all of them are well described by a lognormal distribution, separately in each band for the Gaussian noise and PSF. Fitting a lognormal distribution to each with scipy.stats.lognorm.fit, we get the following fit parameters (shape $\equiv \sigma$, loc $\equiv \mu$, scale $\equiv \alpha$), which can be used to get the probability density through
\begin{equation}
    \label{eq:lognorm}
    p(x) = \frac{1}{(x-\mu) \sqrt{2\pi \sigma^2}} \mathrm{exp}\left( -\frac{\mathrm{log}((x-\mu)/\alpha)^2}{2 \sigma^2}\right),
\end{equation}
and we show the best-fit parameters in Table~\ref{tab:augmentations}.

Each image in the dataset already has a certain level of Gaussian noise and Gaussian blur applied during the observation process, so to add noise on top of what is already in the image we do not want to directly sample from the fit distributions. Instead from the powerlaw distributions we sample an initial noise level in the image,  $\sigma_i$, and a desired noise level in the image,  $\sigma_f$, and apply a differential noise, $\sigma = \sqrt{\sigma_f^2 - \sigma_i^2}$. If  $\sigma_f$ is smaller than  $\sigma_i$ we do not apply the augmentation for that sample in that band.

\begin{table}
\centering
\begin{tabular}{c|l|l|l}
\textbf{Augmentation} & \textbf{shape} & \textbf{loc} & \textbf{scale}\\ 
\hline
Gaussian Noise & (0.2264926, 0.2431146, 0.1334844) & (-0.0006735, -0.0023663, -0.0143416) & (0.0037602, 0.0067417, 0.0260779) \\
PSF & (0.2109966, 0.3008485, 0.3471172) & (1.0807153, 1.2394326, 1.1928363) & (1.3153171, 0.9164757, 0.8233702) \\
Extinction & 0.67306 & 0.001146 & 0.03338
\end{tabular}
\caption{Measured lognormal fit parameters to the data distributions for each of the (g, r, z) bands. See Equation~\ref{eq:lognorm}.}
\label{tab:augmentations}
\end{table}

\bibliography{bibliography}{}
\bibliographystyle{aasjournal}

\end{document}